\documentclass[conference]{IEEEtran}
\IEEEoverridecommandlockouts
\usepackage{cite}
\usepackage{amsmath,amssymb,amsfonts}
\usepackage{algorithmic}
\usepackage{graphicx}
\usepackage{textcomp}
\usepackage{xcolor}
\usepackage{longtable}
\usepackage{pdflscape}
\usepackage{url}
\usepackage{supertabular,booktabs}

\def\BibTeX{{\rm B\kern-.05em{\sc i\kern-.025em b}\kern-.08em
    T\kern-.1667em\lower.7ex\hbox{E}\kern-.125emX}}
\begin{document}

\title{Multi-Modal Data Fusion in Enhancing Human-Machine Interaction for Industry 4.0 Robotic Applications: A Survey\\

}

\makeatletter
\newcommand{\linebreakand}{%
  \end{@IEEEauthorhalign}
  \hfill\mbox{}\par
  \mbox{}\hfill\begin{@IEEEauthorhalign}
}
\makeatother

\author{\IEEEauthorblockN{Tauheed Khan Mohd}
\IEEEauthorblockA{\textit{Dept of Math and Computer Science} \\
\textit{Augustana College}\\
Rock Island, Illinois \\
tauheedkhanmohd@augustana.edu}

\and
\IEEEauthorblockN{Nicole Nguyen}
\IEEEauthorblockA{\textit{Dept of Math and Computer Science} \\
\textit{Augustana College}\\
Rock Island, Illinois \\
nicolenguyen19@augustana.edu}

\and
\IEEEauthorblockN{Ahmad Y Javaid}
\IEEEauthorblockA{\textit{EECS Department} \\
\textit{The University of Toledo}\\
Toledo, Ohio, USA \\
ahmad.javaid@utoledo.edu}

\linebreakand
\IEEEauthorblockN{Sarfaraz Masood}
\IEEEauthorblockA{\textit{Dept of Computer Engineering} \\
\textit{Jamia Millia Islamia}\\
New Delhi, India \\
smasood@jmi.ac.in}
}


\maketitle


\begin{abstract}
Human-machine interaction has been around for several decades now, with new applications emerging every day. One of the major goals that remain to be achieved is designing an interaction similar to how a human interacts with another human. Therefore, there is a need to develop interactive systems that could replicate a more realistic and easier human-machine interaction.On the other hand, developers and researchers need to be aware of cutting-edge techniques being used to achieve this goal. These systems can be combined with Artificial intelligence to make accurate actions or decision. Systems such motion tracker, Virtual Reality headset all make use of AI to reduce error margin and also get the best output from the device. Having a system which is able to not only take input from users but understands this data takes human-machine interaction to the next level.  We present this survey to provide researchers with state-of-the-art data fusion technologies implemented using multiple inputs to accomplish a task in the robotic application domain used in applications for Industry 4.0. Moreover, the input data modalities are broadly classified into uni-modal and multi-modal systems and their application in myriad industries, including the health care industry, which contributes to the medical industry's future development. It will help the professionals to examine patients using different modalities. The multi-modal systems are differentiated by a combination of inputs used as a single input, e.g., gestures, voice, sensor, and haptic feedback. All these inputs may or may not be fused, which provides another classification of multi-modal systems. The survey concludes with a summary of technologies in use for multi-modal systems.
\end{abstract}


\section{Introduction}

Human-machine interaction (HMI) seeks to enable machine adaptation to the application needs and involves the study of interfaces that facilitate the synergy between people and a machine. Also known as human-computer interaction (HCI), HMI primarily classifies interactions as unimodal and multimodal based on the number of modalities used in communication for interaction. Unimodal interface examples include a touch, speech, or gesture-based user interface \cite{adkar2013unimodal}.  Multimodal HMI combines natural input methods such as speech, touch, gestures, pen, or body movements in a synchronized manner and then transforming it to a multimedia system output. These system make use of novel technologies for drawing I/O and allow users to complete out activities with higher accuracy and precision despite any physical constraints or limitations \cite{oviatt1999ten}. The application of the multimodal systems is widely popular in robotics, fixed or mobile, and serves as one of the primary motivations for research in this field with the goal of developing a user-friendly, highly precise system. So far, the fixed robots used in industry are either used through a control stick or keyboard while mobile robotics mostly are autonomous. Many warehouses, such as Amazon and Walmart, have already employed mobile robots to manage the storage and retrieval of items for faster and efficient deliveries of packages and delivery services. Though these robots are not all-powerful and still need humans to work with them, having a multi-modal interactive capability may enhance the operational efficiency of these and similar industrial applications.

While individual modalities may be used in case a user prefers one way of interaction than the other, traditionally, data fusion is employed to create a multimodal system through the fusion of several modalities by allowing the system to decide the accuracy of modality and combined them for interaction. Multimodal data fusion (MMDF) integrates the input of different modalities to enhance the strengths and reduce the deficiencies of the individual inputs. MMDF engines are used to perform these integrations and are used to interpret the various data streams which have different applications in different scenarios depending on the user, time, context, and task \cite{lalanne2009fusion}.  MMDF can be accomplished at the sensor, feature, decision or matching-score level. Sensor fusion is the combination of input streams coming directly from sensors such as webcams or microphones to form a composite  for the applications used in Industry 4.0. These systems are trend towards fully automated and functional cyber-physical systems connected to the internet. In feature-level fusion, the features of individual modalities are combined before the decision is made while decision-level fusion involves combination after the individual decision of each modality is available, and score-level fusion generates a mean score of all modalities based on the individual scores of each input before making a decision. 

In the twenty-first century, multimodality is gaining a lot of attention owing to its advantages over unimodality, such as making use of more senses that connects machines more naturally to the humans, allowing new ways of robust interaction that are fast and efficient and allow respective disambiguation of identification errors \cite{keates1998use}. The primary motive is to enable systems to work either standalone or fused with other modalities depending upon user accessibility. Multimodal interaction is a normal human being life; we talk, use gestures, move around, and shift our gaze for an effective flow of communication. A good example is the use of Google Maps application while driving where the user cannot type the address but can input using speech. Moreover, the application is robust enough to accept input through both modalities \cite{zhang2011traveler}. The research done in these areas requires very robust and precise system architecture and hardware, which should remain relevant for several years. Although multimodal systems are supposed to be user-friendly, newly developed systems might require user training to allow the users to become proficient. The interaction with any system requires a significant cognitive load to understand the system, and there is a risk of making errors that can break the system at any point. The system\textquotesingle s robustness could be increased by employing several mechanisms, including but not limited to training, fault tolerance, data storage scalability, sophisticated UI, logging, and learner mode for new users.

This work is a comparative study where the primary motive is to analyze the state-of-the-art progress in the field and note its accomplishments \& gaps to extend and implement it in broader domains. Moreover, the survey also attempts to explore multimodal applications categorized by a combination of modalities used. The remainder of the paper is structured as follows:. Section 2 and 3 presents a detailed discussion on unimodal and multimodal systems along with their applications, i.e., HMI using voice, gestures, haptic feedback, and their combinations, specifically focusing on modalities used, types of fusion, and their accuracy and precision. In section 4, various MMDF techniques and their applications are explored. We have analyzed various robotic applications including industrial automation and handheld devices. The paper concludes with the analysis of multimodal systems and their pros and cons in terms of reliability, usage, and performance.

\section{Unimodal Systems}

Unimodal systems are designed with single-channel input and are thus confined to a single mode of HMI \cite{karray2008human}. Three popular broad categories of unimodal systems are based on haptic, gesture, and speech. Common examples of interfaces used in such systems are textual, graphical/video and touch. Unimodal systems implemented in the 1990s were widely used in the automation and health care industries. The unimodal systems used in surgery during the early 1990s \cite{mavridis2015review} and are capable enough to understand pre-recorded voice commands and accept only single modality, i.e., speech. They possess many similar limitations; among them a prominent one is accepting pre-defined \textquotesingle canned\textquotesingle \space inputs and severely lack in accepting dynamic human speech. Some example robots included MAIA \cite{mavridis2015review}, RHINO, and AESOP.

In most cases, the human has to initiate the dialog; the systems do not support the flexibly mixed initiative. The robot is incapable of locating the robot physically and unable to respond with their location coordinates; in a few cases, the robots support a ‘canned’ feature. Robots are unable to handle effective speech; that is emotions are neither perceived nor produced. These robots are only capable of handling speech with only a few pre-defined commands. Their non-verbal communication abilities do not exist; for instance, gestures, gait, facial expressions, and head nods are neither perceived nor created. The robots do not support machine learning; they do not learn from the data provided or generated by them. \cite{mavridis2015review}.

Unimodal systems are broadly classified into four general categories, which are haptic, gesture, visual and speech. In our analysis, we investigate each of them. The conceivable strategies utilized for visual are face location, gaze, facial expression, lip-reading, face-based identity, and other client attributes, for example, age, sex, race, and so forth, while voice is actualized through speech input. The other input strategies, for example, haptic and gesture are accomplished through pressure, touch, and nonverbal communication.

\subsection{Haptic Feedback}
Haptic feedback is a field of research exploring human perception and interactions facilitated via the sense of touch, comprising hardware and software able to deliver touch feedback. Haptic communication refers to the use of artificially formed haptic prompts as a medium for communication between two or more individuals \cite{augusto2010ambient}. Multi-touch devices have established acceptability in public spaces, with huge displays appearing in markets, educational institutes, commercial residences, and other areas of high traffic concentration. These systems are designed to adopt changes over time on their interface, allowing users to interact with minimal input from the user \cite{carter2013ultrahaptics}.

This sense of touch via haptic feedback finds application in a variety of consumer handheld devices which are in turn used by a variety of people. The use of haptic feedback and its effectiveness with one particular group, older adults, was explored in 2015 by ECOMODE in Trento, Italy \cite{ferron2015mobile}. The study explored how elderly people interact with portable devices along with how they utilize various applications based on mid-air input interaction. The fourth generation of hand-held devices, which comprises cell phones, tablets, and PDAs, heavily relies on touch input. Various input methods are used by these devices which includes pinching, swiping and double click which is not familiar with older adults. In the initial survey performed by ECOMODE team \cite{ferron2015mobile} on elderly people in order to get their feedback in learning new technologies and challenges faced by them while using mobile devices, the researchers found that the subjects were not always disinterested in the technology, but they were very interested in using the technology to interact with relatives or to get useful information. The challenges faced by the researchers when dealing with the elderly subjects included unfamiliarity with charging ports and fragile ON/OFF buttons. There were also dexterity issues while attempting to use touch gestures with small icons, and users with low vision also made the hand-held devices less user-friendly. 

An experiment was performed using ECOMODE on six elderly people in which they were asked to click on ten photographs using a Samsung Galaxy S5, an iPad mini (eight-inch tablet), and a Samsung Galaxy Tab S 10.5 (ten-inch tablet) \cite{ferron2015mobile}. The feedback received indicated that most users preferred the larger screens of the tablet over the smartphone. The other issues raised by the experimental subjects as follows: 
\begin{enumerate}
\item Lack of clear feedback after clicking the photograph
\item Presence of ambiguous items on the desktop
\item Presence of extra cover/stand on the device
\item Difficulties due to reflections on the screen
\end{enumerate}

There is no empirical data provided by the authors; rather, it is simply a proposal that explores how to make handheld devices more usable for elderly people. 

In an autonomous vehicle, for better control inside a car, haptic features Figure \ref{fig:Haptic} are being added to enhance the feedback a person gets in a vehicle. Bosch showcased gesture control to help control different functions in the car. Ultrasound waves that hit the hand, making it feel as though there is a knob there, but there is no physical knob \cite{Haptic:online}.

\begin{figure}[h!]
            \centering
            \includegraphics[width=0.4\textwidth]{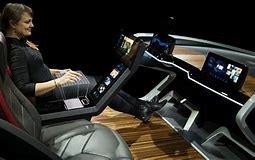}
            \caption{Haptic feedback}
            \label{fig:Haptic}
\end{figure}

\subsection{Gesture Input}
Gestures are expressive and meaningful body motions involving physical movements of the fingers, hands, arms, head, face, or body with the intent of transferring significant information while communicating with the system \cite{mitra2007gesture}. Research has been carried out to improve the accuracy of the gesture input. The researchers have used, a Baxter Research Robot (“Yogi”) and a PR2 (“Kodiak”) \cite{lenz2015deep}. The research deals with improving grasping capability of a robot. In this experiment, they have used Microsoft Kinect on top of Baxter robot angled downwards at roughly 75 degrees toward a table in front of it, which enables the robot to detect things and pick them up from one point to drop them at another. The empirical data shows the results are 84 and 89 percent accurate with Baxter and PR2, respectively. Gestures were used in another implementation called ModDrop, which is based on adaptive gesture recognition. The term "multi-modal" over here refers to various gestures captured through the left and right arm both arms are treated as different modalities. The research focuses on gesture-based detection on multi-scale and multimodal deep learning \cite{neverova2015moddrop}. The researchers have captured spatial information on users to initialize the modalities carefully and to fuse them for cross-modality connections while preserving the uniqueness of every modality.

Ferron et al., (2015) while working with older adults using ECOMODE, finds cell phones can also use mid-air gestures for interaction, interestingly it is much appreciated by them expressing positive comments the possibility of using mid-air gesture \cite{ferron2015mobile}. MYO armband were provided to two elderly ladies, and both of them found this interaction modality interesting. They can control a music application using a MYO armband. Other experiments have been performed that also leveraged gesture input. In one such experiment, the developers attempted to build a healthy relationship between a robot and a human being. Like most of the researchers within the human-robot interaction field, they also utilized the Microsoft Kinect. The initial application for the project is in the homes of seniors where humans aren’t easily available to help the resident. Robots would be deployed to improve the services provided to them. 

The result of the research project was a service robot called Donaxi. The robot has an omnidirectional navigation system \cite{vasquez2015using} with four wheels (each one containing a DC motor and encoder) and a laser system on the front and back for mapping and navigation. The robot is equipped for understanding both voice and motion utilizing the Microsoft Kinect, yet they have utilized motions for the implementation. The development team has participated in the Mexican Robotics Tournament (TMR2015) 
\cite{vasquez2015using}. The Donaxi robot is trained with many videos from different people in different places, which enables the system to work with a variety of users and environments. After providing gesture input to the Microsoft Kinect, the system was trained completely. The above mentioned \ref{table:1} explains that it requires enormous effort to train the system, that is attention is needed for 417 iterations to get trained properly, which is a substantial effort for a single user command. The experimental results show that once the system is functioning, it would able to recognize Attention, Stop and Right for any user as shown in Figure \ref{fig:donaxi}. The work mentioned is not complete, and more features are planned, including the understanding of the additional gestures.

\begin{table}[h!]
\centering
\caption{Total Number of Gestures in the Dataset}
\label{table:1}
\begin{tabular}{|c c c c c c c c|} 
 \hline
 \textbf{Gesture} & Stop & Come & Left &  Right & Attn & Indication & Turn\\ [0.5ex] 
 \hline
 \textbf{Number} & 194 & 177 & 395 & 407 & 417 &  363 & 207\\ 
 \hline
\end{tabular}
\end{table}

\begin{figure}[h!]
            \centering
            \includegraphics[width=0.5\textwidth]{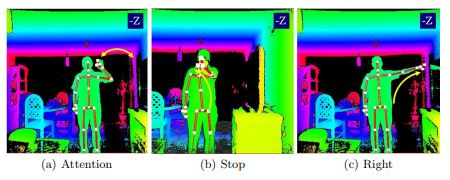}
            \caption{Three Gestures Used at TMR2015 for Donaxi}
            \label{fig:donaxi}
\end{figure}

\subsubsection{Leap Motion}

Leap Motion is a motion control device that connects to a computer and enables users to manipulate objects with their hand motions. The programs are designed to recognize and interpret gesture-based computing to create designs, play games, or carry out some other type of task. Leap Motion is a real-time interaction that can manipulate digital objects. You can run Leap Motion on devices like MAC and Windows. This virtual reality is a great tool of the future, but it is not nearly where gesture control needs to be for the HCI to be smooth. A new update called Orion has provided some recent updates to the fourth-generation core software to improve the finger tracking and motions, faster and more consistent hand initialization, and more accurate shape and scale for the hands just to name a few. The hardware is made with two cameras and three infrared LEDs, which track light with a wavelength of 850 nanometers. The interaction areas are eight feet apart while the motion controllers are 2.6 feet apart after the Orion update. Having gestures like these makes it difficult to interpret or transcribe items like bulbs, daylight, and halogens, which would light up the scene \cite{Leap:online}. The Figure \ref{fig:LeapMotion} shows how the sensory object orientation of the VR works.

A little USB gadget called a Leap Motion controller is plugged into your computer. The Leap Motion controller scans an area of eight cubic feet above the device using LED lights and camera sensors. It tracks both hands and all ten fingers as they travel between you and your computer in the open space. The sophisticated program detects your hands and fingers and converts the data into computer information. The Leap Motion controller is a tiny USB peripheral device meant to be put on a physical tabletop, facing upward, and was invented in 2008. It can also be straddling onto a virtual reality headset. Using two shaded IR cameras and three infrared LEDs, the device observes an unevenly curved area, to a distance of one meter. This is then sent through a USB cable to the host computer, where the Leap Motion software analyzes it. Thousands of devices were first offered to developers interested in designing apps for the gadget by Leap Motion. In July 2013, the Leap Motion controller was released for the first time. Leap Motion issued a major beta upgrade to its core software in February 2016. The program, dubbed Orion, is intended for virtual reality hand tracking. \cite{LeapMotion:online}.

\begin{figure}[h!]
            \centering
            \includegraphics[width=0.5\textwidth]{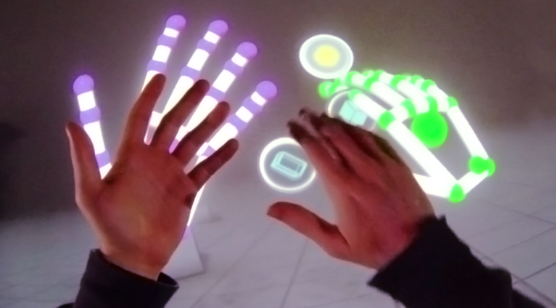}
            \caption{Leap Motion}
            \label{fig:LeapMotion}
\end{figure}

\subsubsection{CaptoGlove}
This glove is a haptic interface system that is Windows compatible and also works with iOS and Android apps. Haptic technology is any which can create a sense of touch by applying force, vibrations, or motions to the user. This Capto Glove Figure \ref{fig:CaptoGlove.jpg} is a motion controller that works using Bluetooth technology. The CaptoGlove claims to be compatible with most VR headset that is already on the market. It is rechargeable, and the battery lasts ten hours. The glove has movement sensors in each finger and a pressure sensor on the thumb. It can be used to play VR, PC, and Phone games or as a controller for many devices and many platforms. You can use just one glove to control and interact, or you can buy both gloves, they cost \$250.00. It claims to be able to control any past, present, or future game created. Most haptic devices are made to interact with virtual reality environments and have sensors that allow you to control and give lifelike feedback. For example, you could probably use the glove to control a car in a racing game, and if you hit a wall or something, it will probably vibrate or shake.

\begin{figure}[h!]
            \centering
            \includegraphics[width=0.5\textwidth]{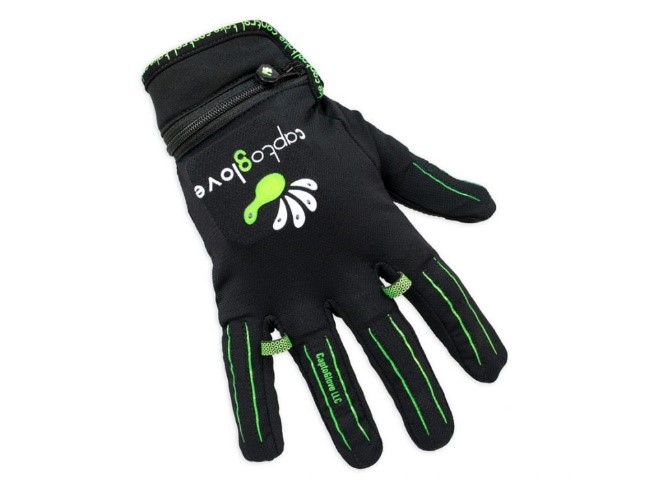}
            \caption{Capto Glove}
            \label{fig:CaptoGlove.jpg}
\end{figure}

\subsubsection{MYO Armband}
The MYO armband is a wearable gesture and motion control device that lets you take control of your phones, computers, and other devices touch-free. Electromyography (EMG) is a technique for evaluating and recording the electrical activity produced by skeletal muscles. The MYO armband allows you to manage your computer, phone, and other digital devices wirelessly via the electrical activity in your muscles. With only a wave of your hand, you can change the way you interact with your digital environment. \cite{MYO:online}. 

\begin{figure}[h!]
            \centering
            \includegraphics[width=0.5\textwidth]{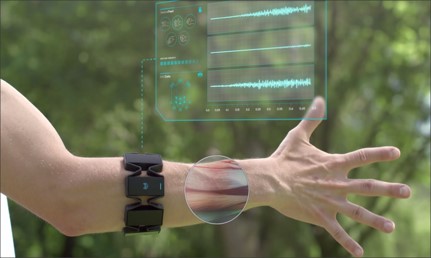}
            \caption{MYO Arm band}
            \label{fig:MYO}
\end{figure}

\subsection{Speech Input }

The last two decades have seen the development of increasing possibilities where computers and handheld devices smart devices can be used \cite{rebman2003speech}. Among them, the feature which was introduced off lately is speech recognition apart from trivial use of keyboard and mouse. Speech recognition is an analysis of the human voice for performing a certain task on a device.  Pursuing a similar concept, a team from MIT is working on enabling human-robot interaction. In this research, a team from MIT and Germany worked together and developed a system capable of understanding and adapting human speech by a robot implemented in a wheelchair \cite{doshi2008spoken}. The system is efficient as it starts the process again whenever it finds any of the following input issues 

\begin{enumerate}
\item User utterance is inconsistent with current discourse (unification with discourse info fails).
\item User utterance can only partly be parsed.
\item User utterance is inconsistent with the robot’s expectations (unexpected info).
\item User asks for the same info several times.
\item No speech can be found in the user utterance.
\end{enumerate}

The system consists of a speech recognizer, a natural language parser, and a dialogue manager. The drawback of this system is that it only accepts speech as an input, while they have claimed it as multi‐modal \cite{doshi2008spoken}. 

A newer feature, created a little less than ten years ago, that has greatly changed our interaction with computers is the voice assistant. Depending on the device or OS, assistants like Siri, Cortana, and Alexa have made interaction with our devices a little easier. They recognize your voice and can be summoned with a simple phrase. The voice recognition software can understand numerous commands and can help you carry them out hands-free. It can start many apps, schedule appointments, return calls or texts, search the internet for the user, and can also give you directions and much more through voice commands. Alexa is set up in your house, and she can play songs for you on demand. She can also place an order through Amazon Prime for you. If you have lights or appliances set up on smart switches, you can ask her to turn them on or off for you. Speech recognition can also help you drive safer because you won’t have to text and drive; you can prompt Google or Siri to compose and send a text to one of your contacts \cite{VoiceChangesEverything:online}.

\subsubsection{Siri}
Siri is a technology that uses voice commands. In everyday life, Siri is used across the world on Apple products. An iPhone user can use Siri for just about anything, as long as your phone is set up to recognize when you say "Hey Siri". When the command is said, the voice recognition program responds with, "What can I help with" as shown in Figure Figure \ref{fig:Siri} The program can also be given the command as soon as you say, Siri, simple commands like play a song, call this person, put this meeting on my schedule, and can even be asked sophisticated questions that will direct you to a site that can help you. Siri is one of the most used virtual assistants out there today, with different languages you can set it too and different accents as well \cite{Siri:online}.

 \begin{figure}[h!]
            \centering
            \includegraphics[width=0.5\textwidth]{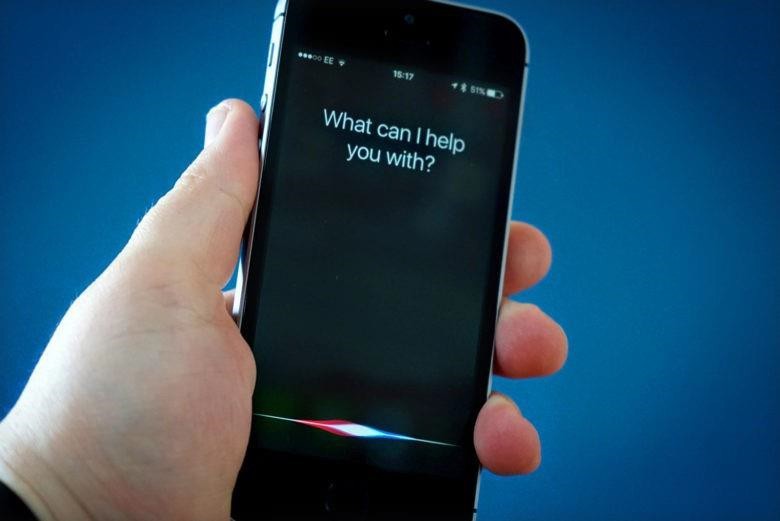}
            \caption{Voice recognition using Siri}
            \label{fig:Siri}
\end{figure}

\subsubsection{Cortana}
Cortana is Microsoft's virtual assistant. Cortana is activated by voice commands and is capable of performing several tasks on your hand-held device. Cortana's biggest market competitor is Apple's Siri. Most users find Cortana very useful for gaming consoles and recommend that everyone take advantage of this helpful tool. 
Cortana can access the internet and do a lot of things that you tell it to. In the past, the users can easily command Cortana to record the gameplay and it is quite easier to tell Cortana to “record that” instead of going to the settings and clicking options, then record, and so on. One cool thing about Cortana is that you do not need a Kinect to make it work, and the player can use a microphone connected to the Xbox controller to make it work \ref{fig:Cortana}. 
Computers also use Cortana, and it is very handy when looking for something on your desktop or even search the web. Cortana is a huge advantage for people that have problems with typing or maybe even some sort of disability. It is still used a lot; though less common in cell phones than video game consoles. The review suggest, more people discusses Siri and use it instead of Cortana. Still, Cortana is used today in several devices, and the research suggest it is doing well because of substantial user base for Microsoft Windows and youngsters who play games using Microsoft Kinect or XBox.
Cortana is also free with an Xbox live membership, but prices vary. One year is \$60, one month is \$10, and three months cost \$25 \cite{Cortana:online}.  
 
  \begin{figure}[h!]
            \centering
            \includegraphics[width=0.5\textwidth]{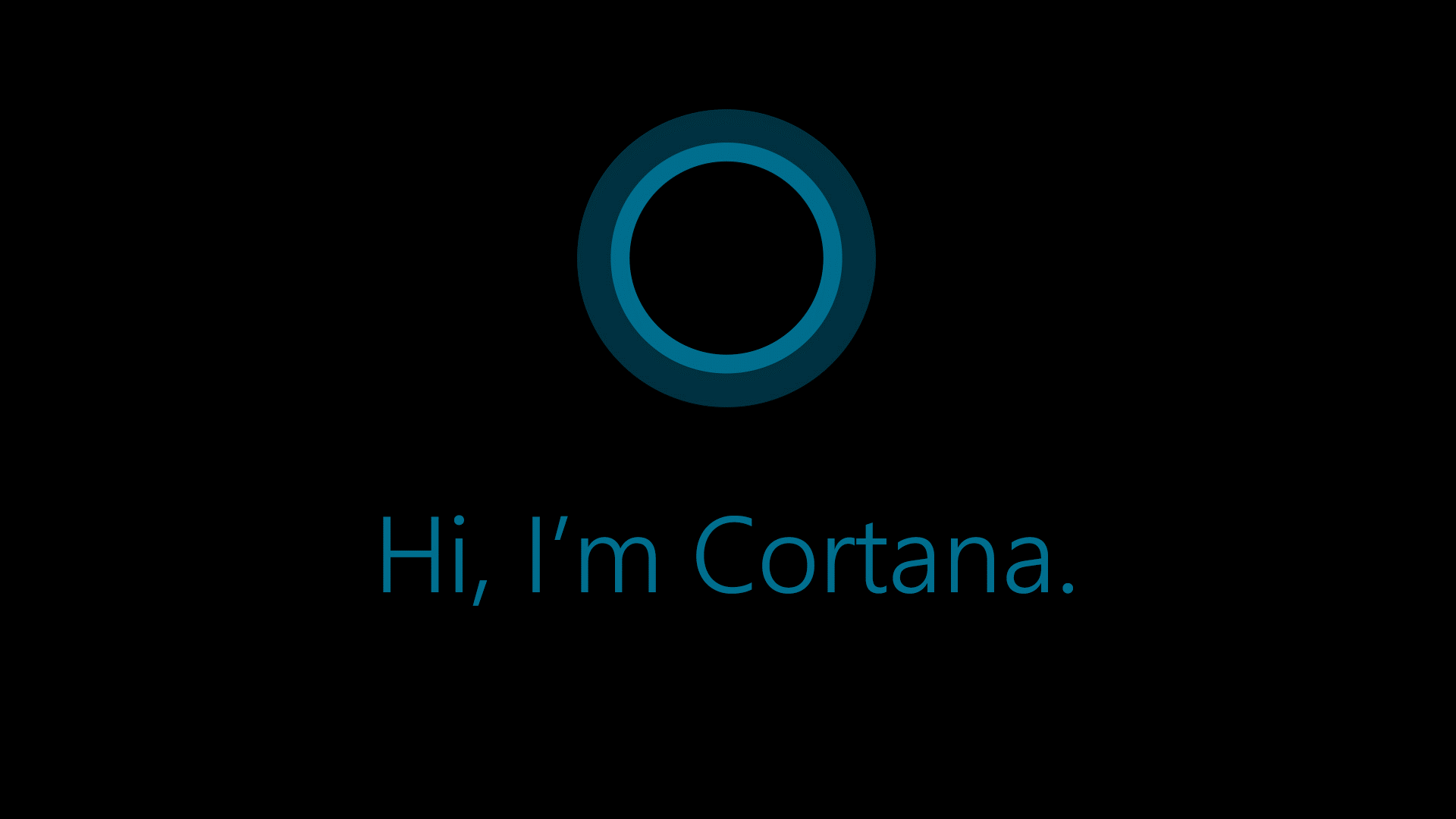}
            \caption{Cortana}
            \label{fig:Cortana}
\end{figure}

\subsubsection{Amazon Echo}

Another popular tool that individuals have been using since 2014 is Amazon Echo. Amazon Echo is a device that is capable of being your personal assistant. It provides information data from the world wide web in real time. It is a conversation voice-control tool that could be used to ask questions, play music, and control other technology devices such as lamps and speakers. Amazon Echo could be integrated into your smart home devices to control the temperature, lock the doors, and also dim the lights. Much like the iPhone tool, called Siri, it can carry on conversations with a person and is constantly ready for someone to talk. To activate it, one must say "Hey Echo" or "Hey Alexa," which will enable the assistant to give voice commands to perform a specific task. The device will start to listen to the area nearby for some type of response to carry out the task. This voice assistant control is known as AVS (Amazon Voice Service), which is an intelligent voice-recognition device that can understand humans. Some recent Echo devices such as the Amazon Look have a built-in camera to take pictures and videos. For example, the sensor is made to learn one's taste by taking photos of different clothes one wears and to have them a better shopping experience using machine learning techniques \cite{Cortana:online}. The Figure \ref{fig:AmazonEcho} shows the layout, and sensory location of the Amazon Echo Look. 

  \begin{figure}[h!]
            \centering
            \includegraphics[width=0.5\textwidth]{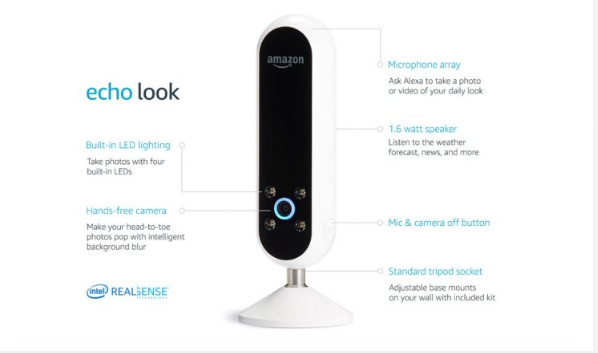}
            \caption{Amazon Echo}
            \label{fig:AmazonEcho}
\end{figure}

\subsection{Eye Gaze Input}
The Eyegaze Edge, created by LC Technologies, is a device that gives users who are not able to use their hands a way to communicate with computer just with the movement of their eyes. The way the technology works is by calibrating the irises with the screen. A small calibration point moves around the screen as the user follows it with his or her eyes and then the eye is entered into the system; this is a quick and easy process. There is a low light, an infrared camera that focuses on the eye and takes you through motions to get a good reading of your eye movement. The Eyegaze Edge Figure \ref{fig:EyeGaze} is built so that the user can calibrate with one eye and navigate with one eye as well. Different means of communication can be used on the system as well including picture icon boards, prestored phrases or store new phrases, computer keyboards for emailing texting or simply taking notes, and even connects to your computer or phone. All this is possible through the image processing software that can determine throgh analysis, where the user's eyes are going to move next \cite{EyeGaze:online}. 
  
 \begin{figure}[h!]
            \centering
            \includegraphics[width=0.5\textwidth]{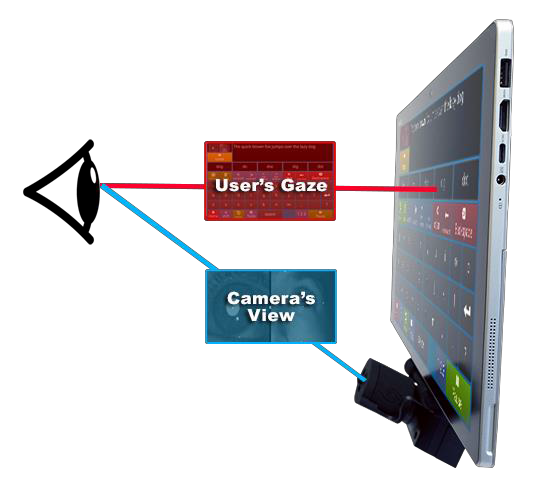}
            \caption{Eye Gaze Input}
            \label{fig:EyeGaze}
\end{figure}

\subsubsection{Tobii EyeTracker 4C}
Eye-tracking technology is a rapidly growing area for both gaming and research, and Tobii is a company that produces many popular devices in this area of technology for both fun and research. Eye tracking Figure \ref{fig:Tobii} is the measurement of the eye's point of sight or motion in relation to the head. Eye-tracking in HCI investigates the scan path for usability purposes or as a method of input in gaze-based interfaces. Although eye-tracking devices can be used for gaming, it seems that the majority of eye-tracking devices are for mainly used for research purposes. There are many types of research being performed with this technology that seem fascinating.  The data from eye tracking devices are being used in psychology and medical research; the data gathered is also being used in marketing research. This particular product is a camera that is \$150.00, and it can perform several tasks, but it also depends on what software you purchase, and they compatible with your device. This camera streams and shows your viewers exactly what you are looking at in real time with ghost software. It is compatible with games available through the same company; it supports more than 150 games. It also has a training tool that is supposed to improve your game-playing skills. It uses a USB connection. It is also a way to securely login to Windows 10. It has some high ratings and good reviews. The main complaints were that it was hard to set up and had hiccups in the games as shown in Fig \ref{fig:Tobii}.

 \begin{figure}[h!]
            \centering
            \includegraphics[width=0.5\textwidth]{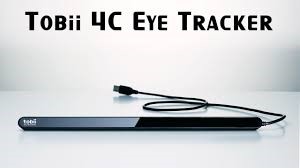}
            \caption{Tobii 4C Eye Tracker}
            \label{fig:Tobii}
\end{figure}

\subsection{Touch Input}
One of the most common modalities we use daily to interact with computers is the touch screen. The majority of people of all ages know how to use the touch screen. After the introduction of the smartphone, along came tablets. Once they became common in most households, then the touch screens became one of the most popular ways to interact with computers. When you use self-checkout at the store, get money from the ATM, place an order at McDonald’s, or rent a Red Box movie, you need to know how to use a touchscreen to complete these tasks. Whether a user likes it or not, CRT and LCD screens are everywhere and have changed the way we interact with computers. Touch screens have given us the ability to use various applications; they are not only used in public information systems such as kiosks at one of the local restaurants or ticketing machines when the user visits the Motor Vehicle Division (MVD) office or any other offices; it is also on devices such as iPhones, tablets, touchscreen TVs, or even touch screen refrigerators, dryer, and washer. As we can observe, the devices are pretty much all around wherever we go, whether it be to wash our cars, to get a parking ticket and find any restaurant; the screen of our PC or even the screen on our vehicles that used for GPS, calling and entertainment purposes or so forth. It is scary to think that our touchscreens are replacing our conventional buttons because of the way they operate. Think about when we had our flip phones and all the buttons that were on them, and after decades, all those buttons have become rare because they are being replaced with the marvelous invention called touchscreens.

There are different types of touch screens; they differ in hardware and software. The ones we encounter when use kiosks at the stores and ATMs use resistive technology Figure \ref{fig:TouchInput}. These screens have two thin layers, one that is resistive and one that is conductive. The screens have a gap in between them with a constant electrical current running through that gap. When you touch the screen, the two screens touch, changing the electrical current, the software read the current change and carries out the instruction related to those coordinates. There are also capacitive touch  screens Figure \ref{fig:CapacitativeTouch}, which are made of materials that hold an electric charge in wires thinner than a hair, arranged in a grid. There are two types of capacitive touch screens; surface capacitive and projected capacitive. They both work similarly; the main difference is that a projective screen has a separate chip for sensing. They work by transferring an electrical charge to your finger, when you touch the screen a circuit is completed, and a voltage drop occurs at that location on the screen. The software then carries out whatever task is related to the location of the voltage drop. Touch screens have completely changed how we interact with computers, making it so those of all ages can interact with a computer and changed the way we complete daily tasks and errands.  There are several devices with finger touch ID (capacitive fingerprint scanner), like computers and phones \cite{TouchScreens:online}. Though a user would think that finger touch ID would use light technology, in reality, the capacitors use electric currents from the spacing the in ridges of your fingers as shown in Figure \ref{fig:TouchInput}. The electricity sends a pulse and gets your print. When the correct fingerprint is read, it will unlock the device, making it much harder for someone to hack you or break into your computer  \cite{TouchScreen:online}.

 \begin{figure}[h!]
            \centering
            \includegraphics[width=0.5\textwidth]{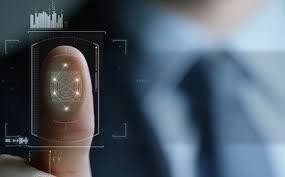}
            \caption{Touch Input}
            \label{fig:TouchInput}
\end{figure}

 \begin{figure}[h!]
            \centering
            \includegraphics[width=0.5\textwidth]{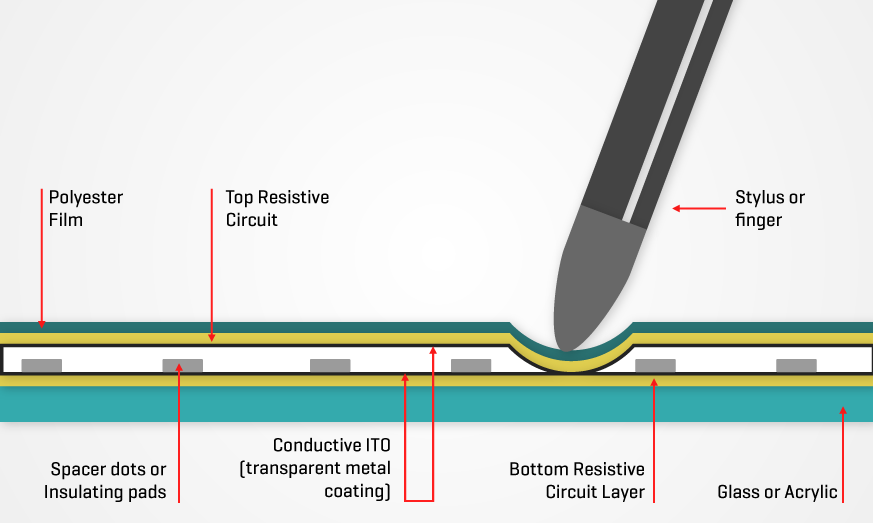}
            \caption{Resistive Touch}
            \label{fig:ResistiveTouch}
\end{figure}

 \begin{figure}[h!]
            \centering
            \includegraphics[width=0.5\textwidth]{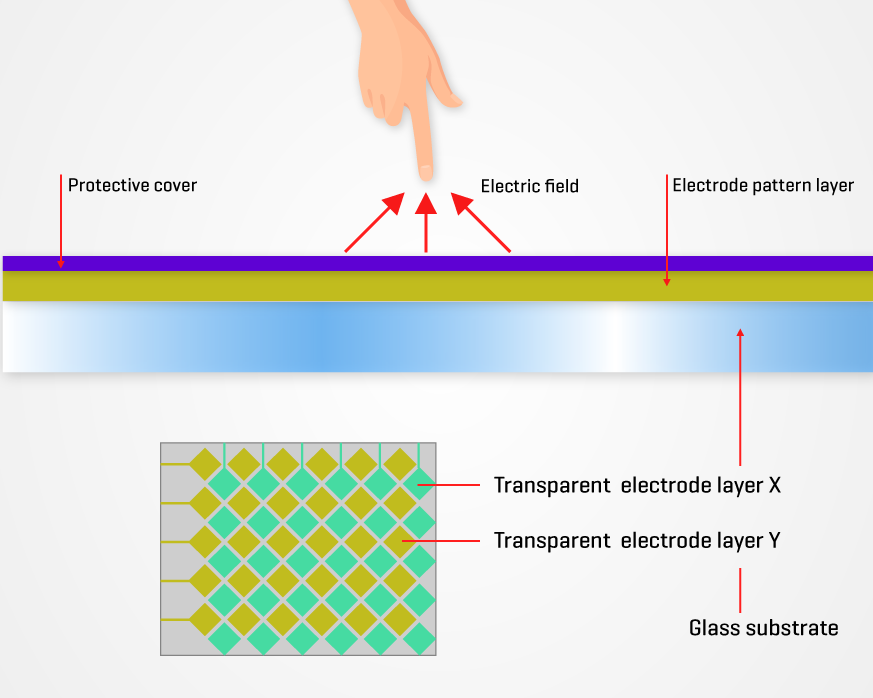}
            \caption{Capacitative Touch}
            \label{fig:CapacitativeTouch}
\end{figure}

\subsubsection{Samsung Foldable Phone}

The Samsung foldable phone is a recent invention that is not yet in the market. It is a touch screen-based devices which accepts input on both sides. The phone was introduced early in 2019 as the first real functioning foldable phone. This dual-battery phone is the first of its kind that brings back the foldable slick aspect of older-generation phones. The phone is one of the most expensive phones in recent times, costing \$1,980 to \$2,600. The Galaxy Fold is made with two foldable screens. The display is made with an ultrathin polymer that uses a new adhesive made by Samsung which enables it to fold many times called an Infinity Flex Display \cite{Galaxy:online}. The figure shows the entire device with the two screens capable of folding as shown in Figure \ref{fig:SamsungFoldablePhone}.

\begin{figure}[h!]
            \centering
            \includegraphics[width=0.5\textwidth]{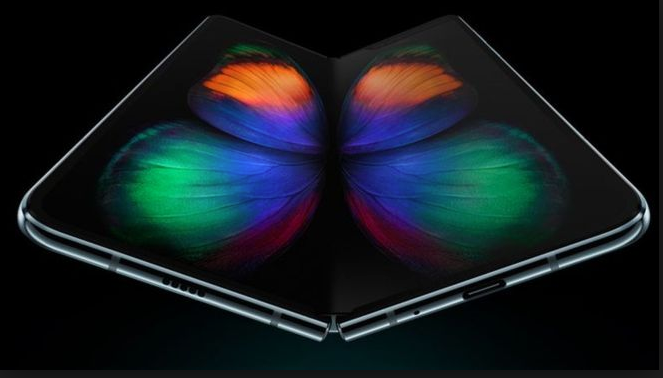}
            \caption{Samsung Foldable Phone}
            \label{fig:SamsungFoldablePhone}
\end{figure}

\subsection{Game Controller Input}
Video games play an integral role in youngsters, lives these days. They spend a daily couple of hours each day on a network-based gaming environment with their classmates and friends, playing games from a remote location. 

\subsubsection{DualShock 4}
DualShock 4 as shown in Figure \ref{fig:DualShock} is a new hardware that supports a PlayStation 4 controller. It helps the user to  navigate through the system and interact with the game and even online with other users. This control has many different features like a touchpad that makes it easy to navigate on the internet like a computer touchpad Figure \ref{fig:DualShock} . It has vibrations that help players in a game know they are in danger; there is a light bar in the front indicating that the controller is on. This controller is also wireless and rechargeable, just like a cellphone. 

 \begin{figure}[h!]
            \centering
            \includegraphics[width=0.5\textwidth]{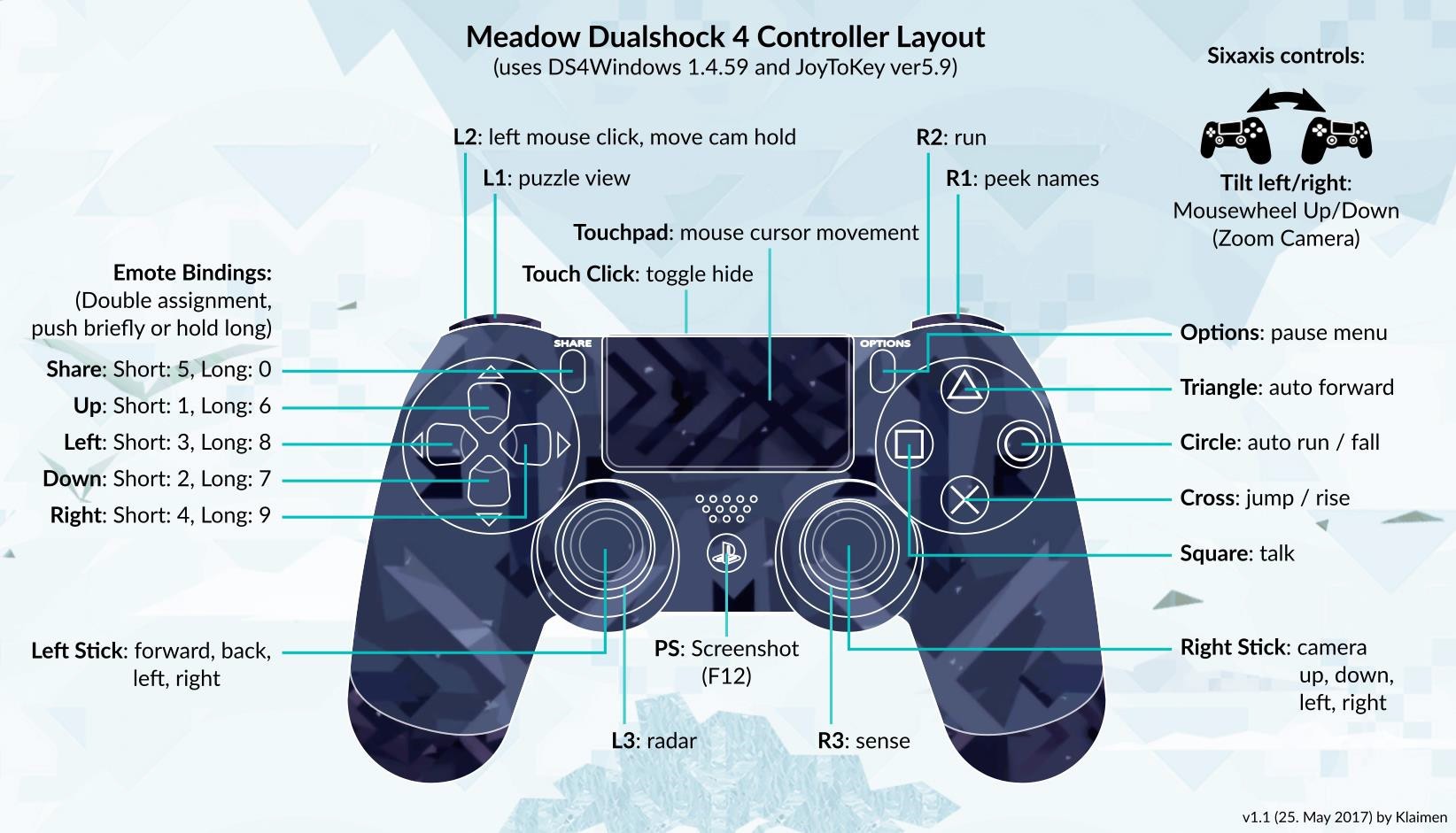}
            \caption{Meadow Dualshock 4 Controller Layout}
            \label{fig:DualShock}
\end{figure}

\subsubsection{Nintendo Switch}Nintendo Switch as shown in Figure \ref{fig:Nintendo} is a gaming console that was released in 2017, although Nintendo is the oldest market player for gaming consoles. The Nintendo Switch is a very successful console because of the classic games that the old-school Nintendo came with.  You can use the switch as a remote control for your tv to turn it off and on and to switch the input on your tv. Some older TV’s might not be compatible. You can find a lost controller by using another controller. The missing controller will vibrate until found. The Nintendo Switch comes with four controllers, and these controllers use different things to control them. They have buttons and triggers, but they can also be controlled by shaking them, twisting them, blowing into a sensor that detects, air and many different kinds of motions. 	
For example, you could be playing a driving game, and you have to hold the controller like a steering wheel and make it turn, just like driving a car. The Nintendo Switch can also be a hand-held device by plugging two of the controllers into the sides of a small handheld monitor. This is the only console that you can really take on the go and also use on a big TV easily. 
Prices vary anywhere from \$299 to about \$500 for a Nintendo Switch. Gaming consoles from Nintendo are worth it because you can have a good time and the cool features. They are really cool user friendly interfaces that you can move the controllers around and perform several inputs like you would in real life \cite{Nintendo:online}.

 \begin{figure}[h!]
            \centering
            \includegraphics[width=0.5\textwidth]{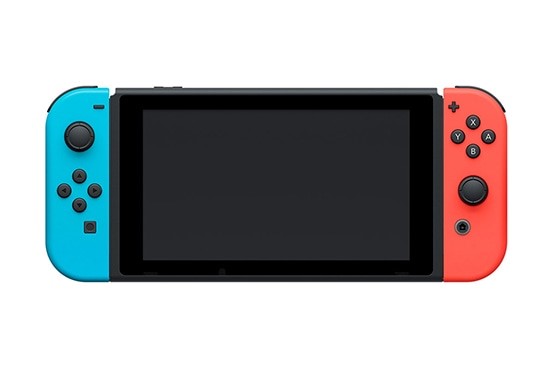}
            \caption{Nintendo}
            \label{fig:Nintendo}
\end{figure}

\subsection{Digital Pen}
Digital pens can transmit your writing to the computer using wireless technology. The pen is thick and packed with digital circuits; just like a mouse, it uses a photocell light detector and a LED light emitter as shown in Figure \ref{fig:DigitalPen}. The difference between the two is that they are stacked vertically instead of horizontally. Also, the pen keeps track of your movements and patterns. Digital pens can upload your writing onto computers though connecting a cord, connecting to a charging dock, o,r best of all, through Bluetooth or infrared \cite{DigitalPens:online}.

 \begin{figure}[h!]
            \centering
            \includegraphics[width=0.5\textwidth]{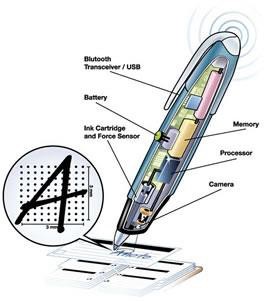}
            \caption{Digital Pen}
            \label{fig:DigitalPen}
\end{figure}

\subsection{Facial Recognition}
Facial recognition is a booming technology for authentication of users from unlocking your computer to finding a criminal or even for finding a friend on Facebook, a Snapchat filter, and so on. Facial recognition takes data and stores it in its system, and when data matches or is very similar, it will recognize the face and bring up the stored face. The police and government agencies use facial recognition when searching for a person. The catch is that the person has to already have a picture in the system, or the system will have nothing to match it to; it would just compare to everyone is in the system to find the best match. Snapchat uses facial recognition to apply filters to faces and can even face swap with other individuals. The same technology is applicable for unlocking computer \ref{fig:FacialRecognition} and even the new cell phones. 

\begin{figure}[h!]
            \centering
            \includegraphics[width=0.5\textwidth]{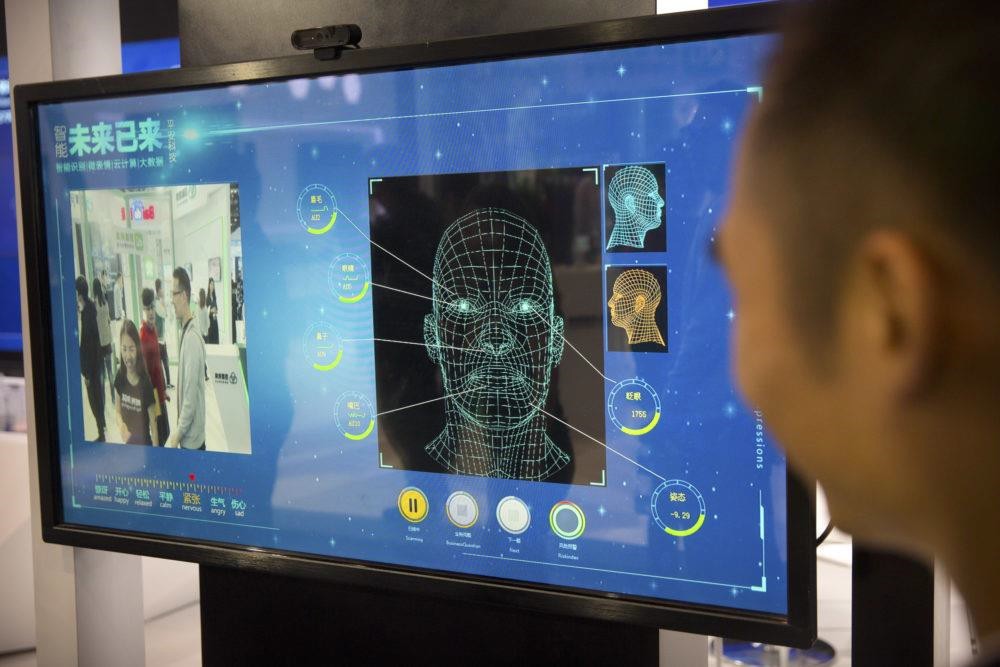}
            \caption{Facial Recognition}
            \label{fig:FacialRecognition}
\end{figure}

\subsection{Drones}
With the recent growth of sensors and distance with network protocols, drones have become a very useful tool to use. Drones are unmanned aircraft that easily be flown remotely or fly autonomously through software-controlled flight plans working along with sensors to avoid obstacles. Drones are used for many tasks, such as search and rescue, wars, surveillance, and fun. Drones are equipped with sensors, cameras, and a software to process the data. These sensors are meant to calculate the distance, time of the flights, chemicals, stabilization, and orientation. Drones are increasingly being used in many different fields. For example, you could use a drone for journalism, disaster response, wildlife monitoring, and agricultural purposes as shown in Figure \ref{fig:Drones}. Drones also have their downfalls as well, especially when it comes to safety and the ethics that are involved with them. drones require special permission to operate, and should not hamper any airspace for airlines and private jets. Recently drones have been on television and criticized for people flying their drones around airports, or also they have been seen in the Olympics. \cite{Drones:online}. 

\begin{figure}[h!]
            \centering
            \includegraphics[width=0.5\textwidth]{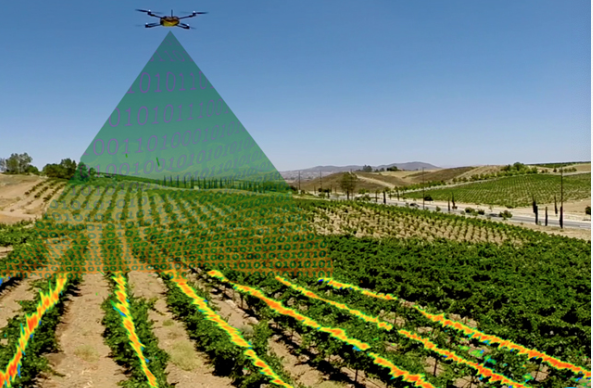}
            \caption{Drones}
            \label{fig:Drones}
\end{figure}

\subsection{EPOC Emotiv headset}
The EMOTIV EPOC+ is a portable, high-resolution, fourteen-channel, EEG system. It was designed to be quick and easy to fit and take measurements in practical research applications. It is compatible with all EMOTIV software products. EPOC emotiv is a brain interfacing devices which accepts input through the USB. It senses the portion of the brain that is active while thinking about specific problems.  \cite{EPOC:online}.

\begin{figure}[h!]
            \centering
            \includegraphics[width=0.5\textwidth]{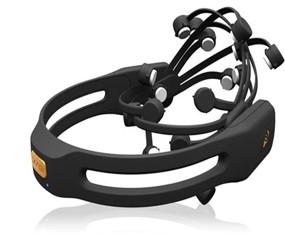}
            \caption{EPOC Emotiv}
            \label{fig:EPOC}
\end{figure}

\section{Multimodal Systems }
Multimodal interaction provides a user with multiple classes or modalities of interaction \cite{oviatt2003multimodal}. A simple example is defined by Bolt’s "Put That There" \cite{bolt1980put} which combines speech with a gesture. Multimodal systems allow the use of multiple human sensory modalities and combine many simultaneous inputs and present the data utilizing a synergistic portrayal of a wide range of output modalities \cite{zaheer2009multimodal}. Multimodality is only truly implemented when there are at least two of the following inputs in any combination: speech, text, or gesture. The usage of multimodalities such as speech, gestures, and haptic feedback can provide a productive, and interface for a media community for various user groups (customary clients, blind users, visually disabled users, and physically impaired users). The investigation recommends the framework does not acknowledge multimodality. The input modalities can be as simple as two pointing devices, or they may include advanced perception techniques like speech recognition and machine vision. The simple cases do not require more processing power than the current graphical user interfaces, but they still provide the user with more degrees of freedom with two continuous input feeds. The multimodal inputs, in this case, are two pointing devices providing input simultaneously to a system. New perception technologies that are based on speech, vision, electrical impulses, or gestures usually require much more processing power than the current user interfaces. The review discusses recently performed research in Multimodal Systems and classifies them based on Fusion.

\subsection{Multimodal Fusion}

Multimodal fusion is the implementation of several modalities in an application that complements the partial input and derives meaningful results. The increasing number of multiple datasets of information, which are obtained by using different acquisition methods through peripheral devices, raised an opportunity to analyze datasets separately and fuse them to derive a common goal. However, until this decade, the data fusion technique is confined within boundaries of psychometrics and chemometrics, the communities in which they evolve. There are several surveys performed by researchers. The first that we explored was published in 2010 \cite{atrey2010multimodal}. The analysis classified the various types of fusion: feature level, decision level, and hybrid level. Methods of multimodal fusion are also discussed, including the rule-based, classification-based, and estimation-based methods. Another study was performed in 2015 that emphasized data fusion, including various methods, challenges, and future prospects \cite{lahat2015multimodal}. In this study, the researchers focused on multimodality in the context of multisensory systems, biomedical systems, and environmental studies. In this study, the goal is to classify the work done after 2015 differently from previous surveys, which is based on classification on the basis of the number of input methods used. 

Recent advancements in technology, including a growing number of domains, leads to an increased interest in exploiting multimodal fusion efficiently \cite{lahat2015multimodal}. Three level of fusion have been discussed, namely feature-level fusion comprising visual features, text features, audio features, motion features, and metadata. The other two levels of fusion are Decision level multimodal fusion \& Hybrid multimodal fusion \cite{atrey2010multimodal}. Decision-level fusion is based on assigning priority to each modality based on its previous experience. Hybrid multimodal fusion is a combination of both feature and decision-level strategies. The methods of multimodal fusion are classified as rule-based, classification-based, and estimation-based. The rule-based multimodal fusion is further classified into three categories:
	
\begin{itemize}
\item Linear
\item 	Majority voting and 
\item 	Custom defined rules
\end{itemize}

Linear weighted-based fusion technique is easiest, based on a first-come-and-first serve basis and linearly combining them. Classification-based multimodal fusion works on assigning linear weight to each modality, for example sum and product, MAX, MIN, AND, OR, majority voting. The classification-based multimodal fusion used the Damster-Schafer and Bayesian algorithm. For the third technique, custom-defined rules, in which the input is either customized per the input needed by the system or as per the need, there is no exact algorithm defined to accept the input. Research published in 2002 claims to implement an algorithm that fuses data of a complex system, represented as 
\[x_i  (t)=A_i x_i (t)+B_i u_i (t)+w_i  (t)\]
\[y_i  (t)=C_i x_i (t)+v_i (t), (i=1, 2….,n)\]
where: 
n     = number of subsystems

xi(t) = state of the ith subsystem

ui(t) = control signal on the ith subsystem

yi (t) = output of the ith subsystem

wi(t) = ith subsystem noise

vi (t) = measured noise of the ith subsystem

This algorithm deals with the fusion of data received from two different input sensors \cite{vershinin2002data}. The experimental results shows among the three performed experiments, only one would function while the other two would malfunctions.

The multimodal gesture recognition algorithm presents a new multimodal framework based on a multiple hypotheses fusion scheme. In this context, the multimodal input is provided by multiple users at a time. An increased number of multimodal inputs brings more challenges to this field. The challenges mentioned in this research are the detection of meaningful information in audio and visual signals, extraction of appropriate features, the building of effective classifiers, and the multimodal combination of multiple information sources. The fusion of multiple inputs can be performed at early, late, or intermediate data/feature levels. The fusion is also possible at the stage of the decision after applying independent unimodal models. This research, just like others we have reviewed before, used Microsoft Kinect, which uses color, depth and audio signals captured by the sensor. The most common approach observed in several research papers published from 2014-2016 is that they have used the Microsft Kinect for recognizing voice and gesture. We believe this is due to the technology's accuracy and precise sensors. The framework talks about accepting multiple gestures as shown in Figure \ref{fig:Proposed_Multimodal_Fusion} from different users and then choosing the best multi-stream hypotheses (input) as shown in Figure \ref{fig:SampleCues} . It involves multimodal scoring and resorting of hypotheses algorithm to manipulate the best gesture; once the algorithm provides the output, the gestures are performed on the system \cite{pitsikalis2015multimodal}.
\[v_i= \sum m \epsilon S  w_m v_m,_i{^s} \]

i = 1 to L
where: weights = w(m) are determined experimentally  

v(m,i)\^s    = standardized version of modality scores based on Viterbi decoding.

Inputs with the highest score are selected for the next phase of the algorithm, which is called Parallel Segmental fusion. The segmental parallel fusion algorithm exploits the modality-specific time boundaries, and it observes the pattern of the sequence of input as occurred in the previous iteration. It was observed during the experiment that there is no one-to-one correspondence between segments, and they are first aligned using dynamic programming. The experimental results presented in this experiment are 93 percent accurate using Microsoft Kinect.  The social signal interpretation (SSI) framework for multimodal signal processing and recognition in real-time was implemented in 2013 at the Lab for Human Centered Multimedia at Augsburg University, Germany. SSI deals with the idea of intuition in next-generation interfaces in real-time. The research seeks to enhance SSI’s C++ API and provide front-end users with the ability to use text editors backed with an XML interface. Computer interfaces are based on explicit commands, but the wave of one's hand or the tone of one's voice can sometimes convey more than hundreds of such commands \cite{wagner2013social}. These natural inputs are much more capable of informing the system about the user’s intuition. To collect and capture human intuition requires a system that is capable of storing a collection of representative samples. Afterward, the collected data is analyzed by a skilled person who works to classify user interaction. To assemble the components mentioned above, the authors proposed a SSI toolbox. The framework proposed was part of the EU-funded Motebo project, and it dealt specifically with physiological data analysis of diabetes patients in an automotive environment using fusion. The concept is not implemented yet; hence no empirical data was provided. 

A survey was performed on Multimodal Interaction at the University of California, Santa Barbara, which discusses the challenges faced while integrating various modalities into a final output and exploring the feasibility of stage, namely, whether it should be done in early versus late integration. Richard Bolt's “Put That There” is one of the most significant demonstrations of multimodal systems. The MIT Architecture Lab (later to become MIT Media Lab) integrated voice and gesture to present geospatial data to a user sitting in a chair. Phrases that the geospatial system understands include the following: "create a blue square there", "move that to the right of the green square" by implementing Multimodal System, "make that smaller", and the canonical "put that there". Early multimodal systems were thus focused on geospatial applications. 

In 1989, another system named CUBRICON \cite{turk2014multimodal} was developed, which enables users to interact using spoken words, gestures, and natural language, and it displayed output in the context of map-based tactical mission planning. Later, in 1993, Koons et al. developed a system that understood speech, gesture, and gaze for a map-based application. In 1997, another system named QuickSet \cite{turk2014multimodal} was designed by Cohen et al., featuring pen and voice-based navigation intended for a US Marine Corps training simulator. In the post-WIMP ("windows, icons, menus, and pointer") period, the multimodal experience was further enhanced to include sketch and 3-D, which was implemented in the “Butler Interface”: interacting with the interface is like interacting with a human who has the ability to speak, gesture, and use facial expressions, among other forms of human communication \cite{turk2014multimodal}, as shown in Figure \ref{fig:Put_That_There}. After 2000, more methods were proposed for interaction that included both verbal and non-verbal communication. Later, the concept of the Perceptual User Interface (PUI) was introduced at a workshop, which eventually grew to be one of the branches of the ACM conference. Input and output modalities, which are relevant in multimodal interaction and fusion, are further classified into modes and channels, as shown in Table 8. Their respective contexts and discrete requirements have distinguished multisensory and multimodal devices.

\begin{figure}[h!]
            \centering
            \includegraphics[width=0.5\textwidth]{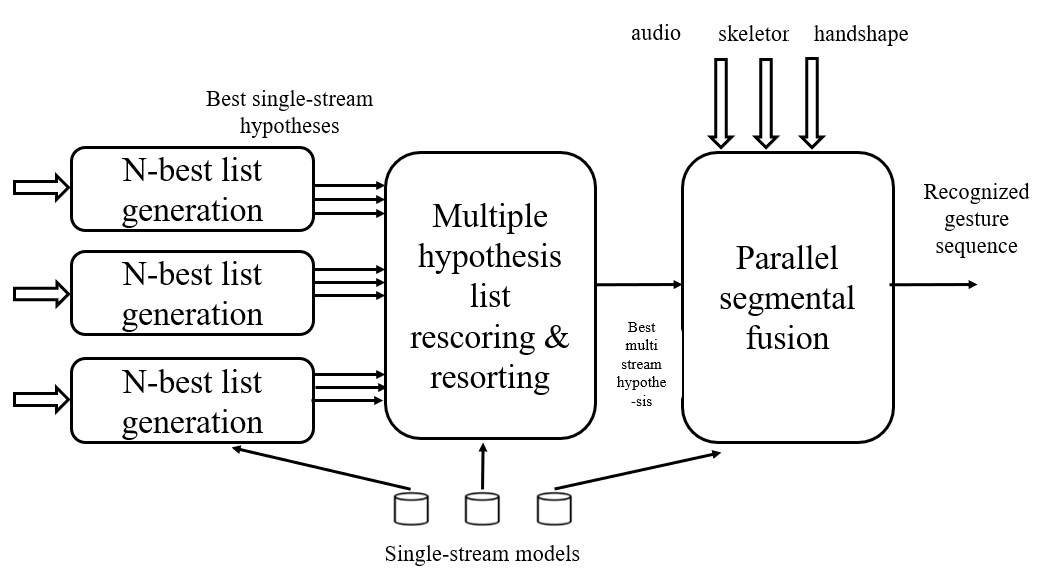}
            \caption{Overview of the Proposed Multimodal Fusion Scheme for Gesture Recognition Based}
            \label{fig:Proposed_Multimodal_Fusion}
\end{figure}

\begin{figure}[h!]
            \centering
            \includegraphics[width=0.5\textwidth]{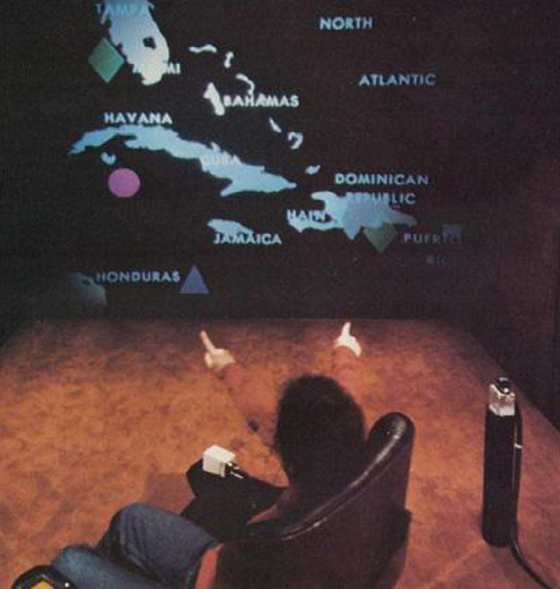}
            \caption{Bolt’s ‘‘Put That There’’ System (Bolt, 1980).}
            \label{fig:Put_That_There}
\end{figure}

\subsection{Multimodal Systems with Fusion }
Speech is a rich channel for human-to-human communication and possibilities to be a rich channel for human-to-machine or computer correspondence. Gestures supplement our speech in various ways, adding redundancy, emphasis, humor, and description, and depiction. Multimodal interfaces made from speech and gesture have more prominent expressive power, adaptability, and convenience \cite{krum2002speech}. With the aim to improve user interfaces of interactive robots with multimodality using speech and gestures, various experiments have been performed. The European Union funds one of the most prominent among them. The research was done by the Comm-Rob project (http://www.commrob.eu) and partially funded by the European Union. This research deals with the fusion of modalities and involves ordering a robot to perform tasks \cite{vallee2009improving}. A Java-based FreeTTS API is used to convert speech into text. The robot used in the experiment is CommRob. The experimental results shown in the article depicts the robot as having very high accuracy. In one of the scenarios, it performed 100 percent accurately when the user called out to the robot robot while gesturing with both hands.  They briefly mentioned that their fusion strategy uses a multimodal grammar that defines which terminal symbols (parts) of a particular command can be provided by which modality. In the following, the researchers present the fusion process with the example of the utterance "Go there [location]." The example chosen is a common phrase used in daily life for a complementary usage of speech (“Go there”) and gesture ([location] indicated by pointing) to produce meaning. They use grammar defined by the following production rules. The generated grammar then produces results by calling a robot function request (goThere(x,y)) \cite{vallee2009improving}.

The command goThere(x,y) explained in above grammar is defined by verb, preposition, and location concatenated together. The resulting meaning is inferred as a command for instance (goThere(x,y)).  The verb is used as "go" with the preposition "there" and location "Gpointing (x,y)" mentioning co-ordinates in the X-Y plane. The research performed using Comm-Rob uses a fusion model available in another research \cite{bogdan2008modeling}  published in Vienna, Austria. In this experiment \cite{bogdan2008modeling}, the authors discussed how to fuse various modalities, but they did not mention explicitly how they  achieved fusion. The speech and gesture input are implemented using natural human conversation, and gestures are considered as good interaction tools. Kendon has defined gestures as voluntary and expressive actions of the human body used together with speech and perceived by the participant as a meaningful part of the speech \cite{kendon2004gesture}. A device extensively used for HMI for voice and gesture input is Microsoft Kinect. The Kinect sensors comprise four different components, which are depth camera, color camera, microphone array, and tilting mechanism. In the research performed in 2013 in Switzerland, researchers fused gesture and voice using Microsoft Kinect. The architecture they utilized includes the Microsoft Kinect API, as shown in Figure \ref{fig:speechgesture}, and captures the input from the user in the form of gesture and speech, which is fused later to draw meaningful operations from it. The results mentioned in the research show the multimodal selection performed better than its unimodal counterpart for total error versus user, namely, a number of errors performed during the experiment \cite{chizari2013combining}. The multimodal system has better error handling capacity. The statistical parameter related to the mean average time of the multimodal selection calculated is 7.5 while the mean average time of the unimodal selection is 16.7, evaluated from Table \ref{table:2}. The results demonstrate that the Multimodal input is better than unimodal regarding the total error.

\begin{figure}[h!]
            \centering
            \includegraphics[width=0.3\textwidth]{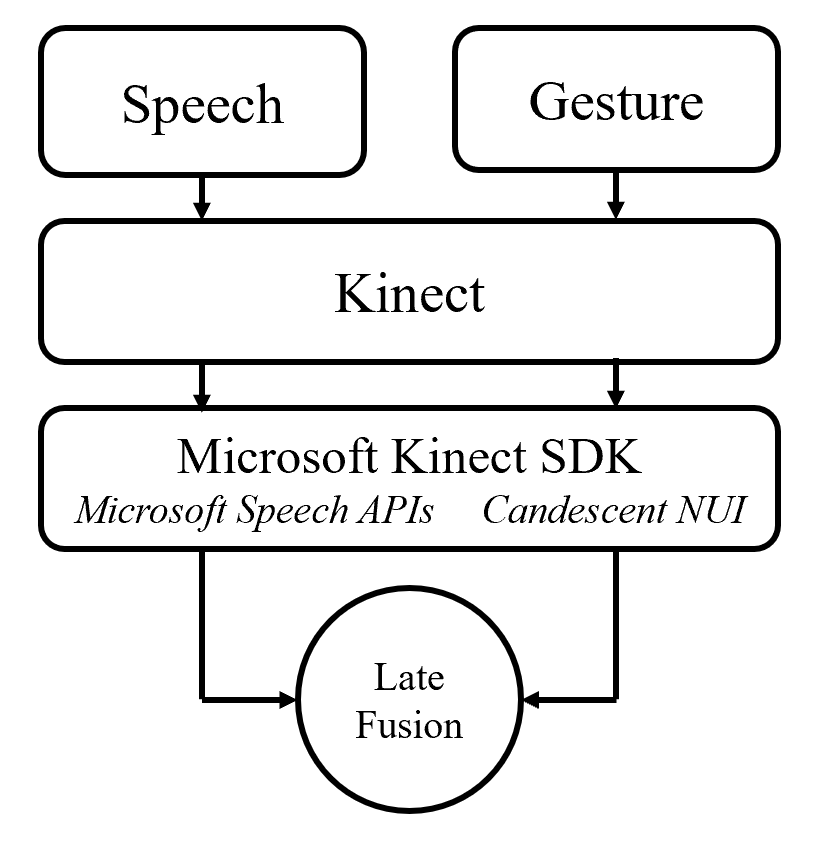}
            \caption{Implementing of Architecture for Integrating Speech and Gesture}
            \label{fig:speechgesture}
\end{figure}

Apart from Microsoft Kinect, to improve accuracy, cell phones can be used for automatic speech recognition and gestures for older adults by using mid-air gesture and voice commands to control the mobile device. For example, Apple's Siri, Microsoft's Cortana, and Google Assistant on Android devices can be fused with mid-air gestures. Though these voice assistants are quite accurate, some advancements should make them reachable to all segments of the society in terms of phonetics, especially Chinese, the Indian sub-continent, and Southeast Asian population, which pronounces various alphabets differently. The speech recognition database does not recognize elderly or shaky voice properly. Consequently, the performance of ASR (automatic speech recognition) reduces 9-13 percent when used by the elderly \cite{ferron2015mobile}. Some possible causes for such a deviation are alterations in the vocal chords, the vocal cavities, and the lungs, along with declining cognitive and perceptual abilities. 

\begin{table}[h!]
\centering
\caption{Comparison of Unimodal and Multimodal Selection: Total Error per User}
\label{table:2}
\begin{tabular}{||c c c c c c c c c c c ||} 
 \hline
 \textbf{Type} & 1 & 2 & 3 & 4 & 5 & 6 & 7 & 8 & 9 & 10\\ [0.5ex] 
 \hline
 \textbf{Multimodal } & 10 & 3 & 13 & 4 & 5 & 8 & 4 & 7 & 3 & 18\\ 
 \hline
 \textbf{Unimodal } & 2 & 3 & 22 & 19 & 4 & 51 & 7 & 5 & 20 & 31\\ 
 \hline
\end{tabular}

\end{table}

ECOMODE is a new generation of low-power multimodal inetrface for cell phones, where the collaboration depends on vocal directions and mid-air gestures. ECOMODE's solution depends on two principle advancements namely mid-air gesture control set and a vision-assisted speech recognition framework. The solution to these issues, proposed by both these modalities are asynchronous in nature. The ECOMODE technology \cite{ferron2015mobile} combines the dynamics of the chin and the motion of the lips to achieve more robustness in the system. The state-of-the-art technology proposed will be designed to work reliably in uncontrolled conditions, particularly under excessive or low lighting and noisy environments.

Multimodal natural user interaction is performed for multiple applications to find the work progress concerning multi-application, multimodal interaction utilizing the Kinect gadget as a two-modular source. The research was carried out in the year 2012. The basic implementation deals with writing an application that understands gesture from Microsoft Kinect, along with consuming input through speech. The system architecture consists of three major components: 
	
\begin{enumerate}
\item Receiving data from different modalities.
\item 	Compiling commands
\item 	Determining and selecting active apps that accept and perform commands \cite{vidakis2012multimodal}.
\end{enumerate}

The application is capable enough of fusing the data and deriving meaning from it. For example, if the system receives a gesture of “Swing Right” from the hand along with a vocal command, then it tries to move the Microsoft PowerPoint presentation to the next slide. Architecture is shown in Figure \ref{fig:gesturevoice} . The advantage of this architecture is that it is not confined to a specific application. It could be used with any application running on a computer. A shortcoming of this research is that empirical data is not provided to support the claim, and there is no mention of accuracy or failure rate while using the specific application. The author concluded by referring to the future design of architecture in such a way that it will support the addition of new modalities, such as tension, pressure, facial expressions, and so forth.

The multimodal systems developed so far have a huge scope of improvement regarding grasping the speech from users of different ethnic groups. Ferreira et al., (2015) proposed a concept of socially-inspired rewards for improving the precision of a system. They are used to quantify reinforcement rewards, which are assigned to users according to how they have interacted with the robots based on vocal interaction. In this method, a potential-based reward-forming strategy instrument is joined with a sample proficient reinforcement learning algorithm to offer a principled structure to adapt to these conceivably chaotic conditions. The experimental setup comprises two live scenarios in which one is responsible for assisting a tourist in a given area called “TownInfo,” and the other is called the “MaRDi dialogue system.” “MaRDi” is responsible for performing a Pick-Place-Carry task in a human-robot interaction context, for instance, “move the blue mug from the living room table to the kitchen table” \cite{ferreira2015reinforcement}. The MaRDi experiment involved a tightly coupled 3-D simulation software known as MORSE, while TownInfo worked as a virtual tour guide. Gesture and voice input were provided and fused to obtain semantics and match with a static knowledge base. 

The recent article published on multilevel sensor fusion with deep learning deals with the fusion of information coming from different sensors. The design was implemented to achieve a trade-off between early and late fusion. At each level of abstraction, the different levels of deep networks are fed to a central neural network, which combines them into common embedding \cite{vielzeuf2019multilevel}. The fusion of audio and visual for evaluating the emotion of a human was performed by Shamim et al (2019), the results achieved are quite exemplary. The experiment receives accuracy of 96.8 percent, and it is fairly good in comparison with other experiments. CK+ database is used, the results achieved through FaceNet2ExpNet is good in comparison with deep sparse autoencoders and DNN using the same database \cite{hossain2019emotion}.

The data fusion using various sensors and audio devices implemented in a hospital room to analyze the environment around the patient and his various needs. This experiment is used for academic purposes and provides detail insights for the College of Nursing students. The various devices used are localization sensors, microphone array, patient simulator, lapel microphones, and physiological wristbands. The experiment is quite a niche in its area, and it’s been implemented quite well for teamwork and its collaboration for a patient \cite{echeverria2019towards}.

The application of data fusion techniques applies to all industries. The same idea is exploited to analyze the learning outcome of a session especially when students are supposed to take classes online. Their activity is observed throughout the class using click-stream data, eye-tracking, electroencephalography (EEG), video, and wristband data. The accuracy achieved was 94 percent using normalized root mean squared error implemented using prediction random forest technique \cite{giannakos2019multimodal}.

The fusion of multimodal data used for rating prediction framework of consumer products by combining EEG signals and sentiment analysis of product review. The experiment uses Emotiv EPOC+ for EEG signal and reviews provided others customer in the form of text. The accuracy achieved is 71 percent. This experiment deals with the crude input provided by Emotiv EPOC and does not get into the minute details of the wavelet \cite{kumar2019fusion}.

Chanoh et al. (2018) implemented a fusion technique called the dense map-centric SLAM method. It is based on combining multiple frames of a handheld LiDAR and compensate the remaining information to complete the image using Fusion. The  Trajectory error in meters achieved using this technique is 0.076m from a length of 360 meter input, and the time frame required is 9.1 minutes. The experiment effectively reduces LiDAR noise by Surfel fusion \cite{park2018elastic}. In a bigger context, if the experiment is performed outside, global optimization may not be achieved.

In another experiment, the fusion of images captured by a camera and GPR/encoder data that are spatially evenly-spaced are captured and fused for subsurface transportation and infrastructure inspection. The proposed algorithms need to improve further for accuracy and speed, but in the ideal condition, the accuracy achieved is 98 percent. The experiment was performed on the bridge deck at the Ernest Langford architecture center at Texas A\&M University to test the system \cite{chou2018encoder}.

Supervised learning is used as training data for a map application while capturing the images from the camera and annotated images. It will input the raw data from user, transmits and apply the algorithm to process the output. The algorithm usually was created by human, and it was used as a function in supervised learning.  An experiment was performed to create inexpensive technology for marking road segmentation. It will help to create rules for maps used in autonomous vehicles. The accuracy achieved in the experiment is 75.04 percent. Moreover, the experiment does not include semantic classification of the road markings to retrieve the rules of the road \cite{bruls2018mark}.

For sensing human reliance on texture recognition, an experiment is performed using GelSight tactile sensor for capturing tactile images, which are further fused with vision using deep neural networks. In this experiment, the accuracy achieved is 90 percent by implementing a fusion method named Deep Maximum Covariance Analysis (DMCA). Using the algorithm based on DMCA, it is easy to calculate the perception performance of either vision or tactile sensing. The limitation of the experiment is temporal information is not included during the experiments \cite{luo2018vitac}.

The application of multimodal fusion was performed in surgery, while the surgeon is using MYO armband for gestures, and EPOC Emotiv for capturing the EEG signals and Microsoft Kinect for speech and capturing the body movements. The accuracy achieved by the experiment is 88 percent. Moreover, the device is too complex; it would not be easy to perform surgery while having two devices on the body. The experiment talks about excluding or minimizing the workforce required in the operation theatre. The idea is to read the surgeon's mind and provide the surgical tools required during the surgery. In a typical scenario, there are around ten professionals required in the operation theatre. If this experiment is implemented in a commercial environment, we will able to reduce the manpower required at the hospital \cite{zhou2018early}.

The application of a robotic arm is used in calculating the depth estimation of metallic pieces. The robotic arm developed consists of Microsoft Kinect, lasers and mono-camera, and the system is designed to find featureless objects such as metallic plates, metallic connectors, and monochromatic objects. After the experiments performed, 95 percent achieved and 100 percent with the help of a human operator. The system is further improved by adding utilities and making it viable to be accepted by the industry \cite{di2018tracking}.

Similarly, the application of multimodality is widely exploited in the health care industry. In an experiment, video and kinematics data used to perform surgical operations. The experiment claims to achieve the accuracy improvement of 15.2 percent using unsupervised trajectory segmentation based on a TCS-K approach. A SCAE network is used visual feature extraction from the input video \cite{shao2018unsupervised}.

The application of multimodal data fusion can be used in capturing cultural attributes, using a sensor simulator and a signal generator. The experiment claims to fuse attributes with heterogeneous information. It can learn new user attributes from distributed data streams such as human behaviors in different situations. The various attributes are talkativeness, extroversion, uncertainty, individualism, and personal distance. The limitation of this experiment is that, it is unable to find information on the user's current state (e.g., mood and satisfaction level, etc) \cite{santos2018extended}.

\begin{figure}[h!]
            \centering
            \includegraphics[width=0.5\textwidth]{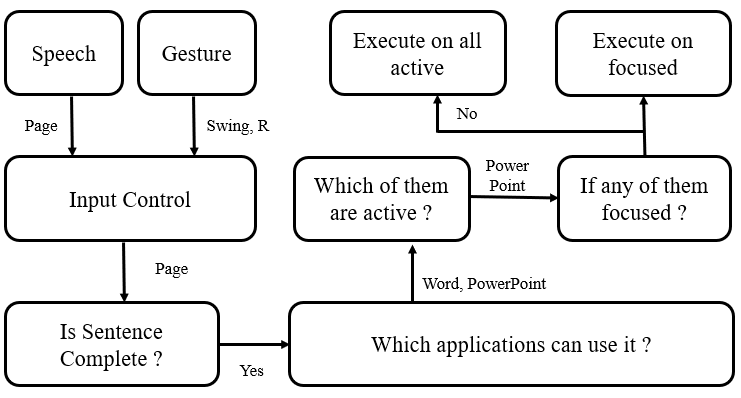}
            \caption{Two Applications Multimodal (Gesture-Voice) Example}
            \label{fig:gesturevoice}
\end{figure}

\subsubsection{Oculus Rift 3D}
The next HCI device that we have explored is Oculus Rift 3-D goggles. The Oculus Rift goggles as shown in Figure \ref{fig:OculusRift3D} are a virtual reality headset developed and manufactured in 2016. The user pulls a helmet over his head, and suddenly, he is inside a virtual world that seems completely lifelike. The user can run around, fight, race, fly, and create ways that gamers (or anyone else, for that matter) have never done before. This is a great form to interact with your devices. It absorbs you into another world, and it makes users interact with the device in ways never thought of. The VR devices are becoming more and more popular as technology is advancing so much. Picture a set of ski goggles, but instead of miles of fresh powder, you are transported into space or underwater. The Rift accomplishes this using a pair of screens that display two images side by side, one for each eye. A set of lenses is placed on top of the panels, focusing and reshaping the picture for each eye and creating a stereoscopic 3-D image. The goggles have embedded sensors that monitor the wearer's head motions and adjust the image accordingly. The latest version of the Oculus Rift is bolstered by an external positional-tracking sensor, which helps track head movements more accurately. The result is the sensation that you are looking around a 3D world. Although it is nice to be able to leave reality for a moment, these goggles are very expensive. However, they are a good way to interact with a device in a virtual transportation to the unknown. This device is capable of having input from several modalities, including gesture and speech, simultaneously \cite{Oculus:online}.

\begin{figure}[h!]
            \centering
            \includegraphics[width=0.5\textwidth]{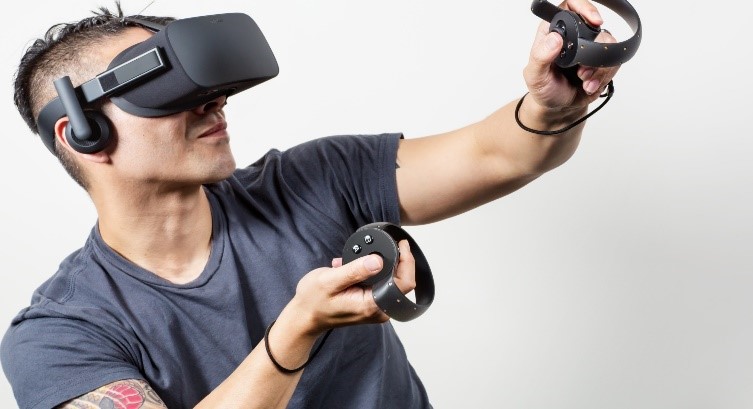}
            \caption{Oculus Rift 3D}
            \label{fig:OculusRift3D}
\end{figure}

\subsection{Multimodal Systems without Fusion }
Multimodal systems without fusion are defined as an application with multiples, but they are not complementing each in the completion of vague of partial inputs. A robot named “Motherese” is used to develop multimodal emotional intelligence. In this study, the authors were attempting to develop the concept of Multimodal 
Emotional Intelligence (MEI) \cite{lim2014mei}. As humans perceive the effect of voice, movement, music, and point light displays, the MEI robot accepts input in the form of voice and maps it to other modalities. The MEI robot uses various parameters to understand and express multimodal emotions that are defined by SIRE (Speed, Intensity, irRegularity, and Extent). The inputs used in implementing the MEI model were .wav audio files, a Microsoft Kinect for capturing gestures, and Flute for the music, as shown in Figure \ref{fig:SIRE}. The generation of emotional expression using MEI is implemented through the intensity and speed of speech. A voice capture with parameters in the ranges could be judged as displaying happiness or sadness accordingly.

\begin{figure}[h!]
            \centering
            \includegraphics[width=0.5\textwidth]{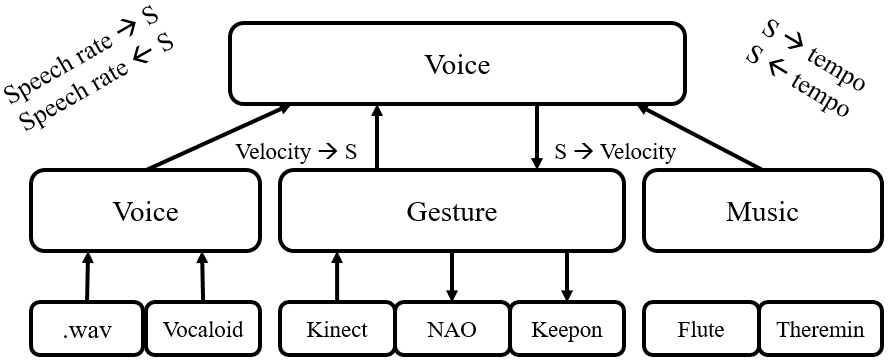}
            \caption{SIRE Paradigm for Experiments across Voice, Gesture and Music}
            \label{fig:SIRE}
\end{figure}

The experimental results show the system is not robust enough to comprehend and analyze all the gestures properly; it sometimes gets confused between anger and happiness mentioned in Table \ref{table:3}; for instance, it has identified Happiness with 62 percent accuracy, whereas it identified sadness, anger, and fear by 0 percent, 19 percent, and 19 percent accuracy respectively. The researchers claimed that the system's accuracy is 63 percent in the first iteration and can be calculated by taking the mean of the diagonal values of Table \ref{table:3}. In the later stages, they  reached accuracy of 72 percent \cite{lim2014mei} which we believe could be improved by training the system in a more diversified environment, such as, a twenty-dimensional confusion matrix. 

\begin{table}[h!]
\centering
\caption{Confusion Matrix}
\label{table:3}
\begin{tabular}{||c c c c c c ||} 
 \hline
 Detected & Happiness & Sadness & Anger  & Fear  & p-value \\ [0.5ex] 
 \hline
 Happiness & 62 & 0 & 19 & 19 & \ <0.0001 \\ 
 \hline
 Sadness & 2 & 90 & 0 & 6 & \ <0.0001 \\ 
 \hline
 Anger & 55 & 0 & 43 & 2 & \ <0.0001 \\ 
 \hline
 Fear & 21 & 12 & 12 & 55 & \ <0.0001 \\ 
 \hline
\end{tabular}
\end{table}

Following a similar concept, a multimodal manipulator control interface was designed which uses speech and multi-mouch mesture recognition. The research deals with managing a robotic arm with touch and gestures \cite{oka2015multimodal}. The degree of freedom for a robotic arm is seven, which is controlled using rotate, open and close commands for the gripper. Per the claim, the robotic arm could be used by novice users, and they could operate the robotic arm easily. The interface recognizes five types of touch gestures: slide, open, close, clockwise, and counter-clockwise. In control mode, open and close are used specifically for the gripper. While the left and right gestures are used to move gripper left and right respectively, the prototype has been developed using the seven degrees-of-freedom robotic arm, using a manipulator, which includes a laptop with a touchscreen \cite{oka2015multimodal}. The robotic arm used to have six joints and a gripper. Figure \ref{fig:Hardware} illustrates the system, comprises of a laptop with a touchScreen, a CAN-to-USB adaptor, a USB headset, and the robot with six joints and a gripper. The touch gestures mentioned in Figure \ref{fig:Hardware} explain how fingers should be used while performing slide, open, close, clockwise and counterclockwise operations to control the iARM. T. Oka et al. (2005) have \cite{oka2015multimodal} implemented the multi-touch gesture recognizer, which detects and understands the gestures on the touchscreen. The system was designed for recognizing a aprticular grammar, which includes three rules and thirty-five words. The interface recognizes multimodal commands using spoken language and only one contact point over the panel. When the system receives a multimodal command, the system sends velocity and position commands to the manipulator, moving it to the defined location using three arms. The pilot study reveals that new users can control the manipulator \cite{oka2015multimodal}. They can easily pick up, rotate, and replace objects using gesture and multimodal commands. The limitation of the results shows that the user would not be able to operate effectively in a rotational mode. 

\begin{figure}[h!]
            \centering
            \includegraphics[width=0.5\textwidth]{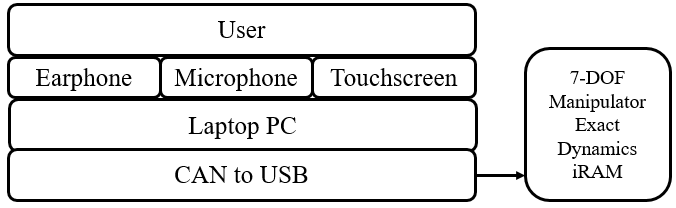}
            \caption{Hardware Configuration}
            \label{fig:Hardware}
\end{figure}

Krum et al. (2002) implemented a multimodal navigation interface \cite{krum2002speech} using speech and gesture for a whole 3-D visualization environment. Virtual Geographic Information System (VGIS) is used as a multimodal interface, which provides a set of 3-D navigation of the globe. Speech and gesture multimodal control is implemented in Earth 3D Visualization Environment; research was done in 2002, which aims at reducing the error rate in multimodality in comparison to unimodal component interfaces. In noisy environments, users can rely on pen and gesture input, while the differently-abled users can use speech. Those users who do not have a clear accent or shaky voice will prefer to use gesture and pen. One of the important aspects of multimodality is that the user may not have peripheral devices, like a mouse or keyboard, to provide the input, and he or she may mostly be occupied using his or her hands for moving around the display most of the time. It is important to understand the limitations of multimodal speech and gesture interfaces rather than comparing performance with other interfaces.
The VGIS systems allow the user to travel from the orbital perspective of the entire globe, which displays 3-D building models and sub-meter resolution images of the earth’s surface. Navigating an extended 3-D space in VGIS brings more challenges to the applications:
\begin{enumerate}
\item Including scale, seven degrees of freedom must be managed. 
\item In a virtual environment, good stereo imagery must be maintained. 
\item Navigation methods must work at all spatial scales.
\end{enumerate}

Krum et al. (2002) were able to address 1 and 3. The implementation part of multimodal interfaces used a variety of hardware in a desktop PC, a laptop, and a Fakespace Virtual Workbench powered by an SGI Onyx2. Voice recognition was performed by IBM ViaVoice in which speech utterances are converted into text and sent as commands over the network; sample commands. A gesture pendant is worn on the human chest which has an LED, and it captures the hand movement. The camera has an infrared filter which is having the best feature; it avoids other light sources from interfering the image. The speech commands are then translated to multimodal interface commands, based on the mapping, which enables the model to render the requested image, building, etc. Navigation commands, as shown in Table \ref{table:5}, are available to navigate in x and y directions. The system works in three modes.

These modes are the orbital mode, walk mode, and fly mode. In walk mode users can constrained to the ground, orbital mode presents a third-person point of view, which always looks from down to above, and the fly mode presents helicopter-like flight. The performance of the system is evaluated on the metrics mentioned below. 
\begin{enumerate}
\item Gesture recognizability and responsiveness: how accurately and quickly the system recognizes gestures and responds
\item Speed: efficient task completion
\item Accuracy: proximity to the desired target
\item Ease of learning: the ability of a novice user to use the technique
\item Ease of use: the cognitive load of the technique from the user’s point of view
\item User comfort: physical discomfort, simulator sickness
\end{enumerate}

\begin{figure}[h!]
            \centering
            \includegraphics[width=0.5\textwidth]{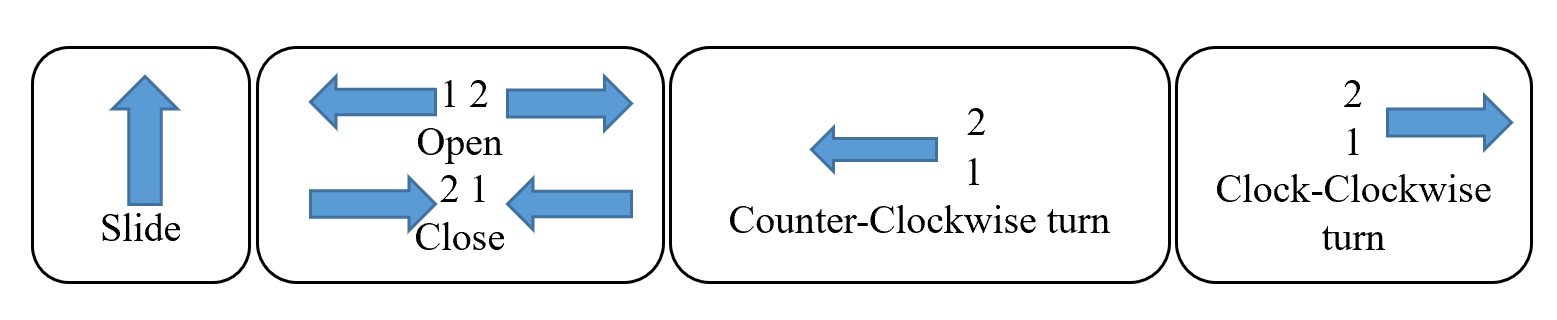}
            \caption{Touch Gestures}
            \label{fig:TouchGestures}
\end{figure}

\begin{table}[h!]
\centering
\caption{Recognized Gestures}
\label{table:5}
\begin{tabular}{|c |} 
 \hline
  Vertical Finger Moving Left: Pan Left\\
 \hline
 Vertical Finger Moving Right: Pan Right\\
\hline
Left Finger Moving Up: Zoom Out\\
\hline
Left Finger Moving Up: Zoom In\\
\hline
Right Finger Moving Up: Pan Up\\
\hline
Right Finger Moving Down: Pan Down\\
\hline
Open Palm: Stop\\
\hline
\end{tabular}
\end{table}

Table \ref{table:6} summarizes both commands. A novice user must go through the training of fifteen to twenty minutes to understand the system and learn commands. Recognizability and responsiveness of voice recognition lags were factors in the performance of the users. Also, users would sometimes have to repeat commands, but overall the system is capable enough to understand most of the times. With regard to speed and accuracy it took a total of 10.1 minutes to complete one task, which is too long, and the user had to utter 50 to 100 spoken commands, whereas the mouse interface just took 3.5 minutes \cite{krum2002speech}. The accuracy of the system is pretty good with a very few chances of error, and that too while adjusting more detailed movements. If we talk about ease of learning, ease of use, and comfort, the system is capable of addressing all these aspects. Some users prefer keystrokes rather speech and gesture. There are certain commands which are not present in the system, or if the user provides a wrong command, the system will prompt "Command not found." The designed system has an ease of use which allows the users to “move higher” command with an increased rate of motion but a decreased rate of motion in case of “lower motion” command, which could be confusing at times because we are addressing both speeds, up and down, with a single gesture, just by altering the speech input. The system is not that comfortable as it fatigues the hands while using the system.

Speech and arm motion were explored in a multimodal context by Bozkurt et al. (2016) which was discovered in Istanbul, Turkey, in 2016. The goal of the research was to implement machine learning in multimodality. In virtual environment designs, gesticulation is an important concept introduced in the paper. Gesticulation deals with adopting and emphasizing the human-centered aspect, which is missing in virtual environments. The study explored a programmed combination of motion in synchrony with speech and joined with nonverbal correspondence segments into virtual character segments. The study deals with the feature extraction of unimodal clustering using both semi-supervised and unsupervised forms of clustering. For semi-supervised learning, a pool of gesture input is provided using the Hidden Markov Model, while in unsupervised learning, a large-scale multimodal dataset is used \cite{bozkurt2016multimodal}.

For the unsupervised learning experiment, the researchers made a twenty-minute recording of five different native users (data are shown in Table \ref{table:7}), all of them being Turkish in origin. The speakers wore a black suit with fifteen color markers and a microphone, placed close to their mouths, and synchronized with their speech. During the experiment, the users were not instructed on how to provide the gesture input specifically. The number of gestures collected by all the users is shown in Table \ref{table:7}. These gestures are analyzed by using the semi-supervised learning technique, which shows that each user performed seven unique gestures.

Further research regarding gesture and voice interaction with interfaces was carried out at the University of Glasgow in 2012 in the research by Rico et al \cite{rico2010gesture} . The paper dealt with the issues of social acceptability and user perception as they related to multimodal mobile interaction techniques. The evaluation of social acceptability explored performance regarding audible and visible interactions, including how the user perceived the interaction and how comfortable they were while using the device. The exact scope and definition of user experience are still debatable, but while designing handheld devices, the designer should take into account the individual thoughts of users and their feelings and reaction to an interface. To understand the user’s behavior, an experiment was carried out with sixteen gestures and sixteen voice commands. Voice and gesture were chosen because of their highly visible and audible nature. The modalities were examined on an individual basis, rather than grouped together. The gesture was classified into four subcategories, namely, emblematic, device based, arbitrary, and body based. The most widely accepted gestures are deemed emblematic, while device-based are those who are involved in directly manipulating a device. Arbitrary gestures are defined as those set of gestures which were defined previously as emblematic and device based. Body-based gestures are those in which direct physical contact with the device is not involved. Body-based gestures work with external sensors and capture the body’s movement. The voice commands used classified into three categories: speech, command, and non-speech. The command is a type of input in which user says short words, for example, “call” or “lock” and so on, while speech input is short commonly used phrases and non-speech input including noises which are used in everyday life (the gesture details are mentioned in Table \ref{table:6}). The experiment was carried out such that half of the users could use gestures while the rest were told to use voice commands. After the experiment, an interview was conducted in which a worksheet was provided to collect feedback on what the users felt regarding input preference, locations where these inputs might be used, tasks where these inputs could be used, and so on. A total of nineteen participants were involved in this study, with the majority ranging in age from eighteen to twenty-nine, while two local community members were between the ages of seventy and ninety-five. The results show that device-based gestures were preferred over other gestures, body-based gestures ranked second, arbitrary gestures were least acceptable or preferred, and emblematic gestures were second least acceptable. In speech input, “commands” are most widely used and had high acceptability in comparison with speech and non-speech commands. Other research, this time focusing not on user interaction with multimodal systems but on how multimodal systems interact with desktop applications, was performed in 2013 \cite{vidakis2013multimodal}. 

\begin{table}[h!]
\centering
\caption{Gestures and Voice Commands, by Category}
\label{table:6}
\begin{tabular}{||c c c c  ||} 
 \hline
  \textbf{Gesture} & \textbf{Category} & \textbf{Voice} & \textbf{Category}\\
 \hline
 OK Gesture & Emblematic & Say "Close" & Command\\
\hline
Money Gesture	& Emblematic	& Say "Open" &	Command\\
\hline
Peace Sign	& Emblematic	& Say "Call"	& Command\\
\hline
Shrugging	& Emblematic	& Say "Lock"	& Command\\
\hline
Device Stroke &	Device-Based &	Say "I'm Fine"	& Voice\\
\hline
Device Shaking	& Device-Based	& Say "Bad Weather"	& Voice\\
\hline
Device Flick	& Device-Based	& Say "That's Nice"	& Voice\\
\hline
Device Rotation	& Device-Based	& Say "So Busy"	& Voice\\
\hline
Upright Fist	& Arbitrary	& Humming &	Non-speech\\
\hline
Hook Finger	& Arbitrary	& Buzzing 	& Non-speech\\
\hline
Sideways Fist	& Arbitrary	 & Say "Chh"	& Non-speech\\
\hline
Open Palm	& Arbitrary	& Doo Doo Doo	& Non-speech\\
\hline
Shoulder Rotation	& Body-Based &	Say "Psst" &	Non-speech\\
\hline
Wrist Rotation &	Body-Based	& Whistling	& Non-speech\\
\hline
Foot Tapping & 	Body-Based	& Clicking	& Non-speech\\
\hline
\end{tabular}
\end{table}

Continuation of a part of previous work, reported by Niloas et al. (2012), the research behind the 2013 paper, was performed by a joint group of the Applied Informatics and Multimedia Department, Greece, the Electronic and Computer Engineering Department of Brunel University, UK \& Medialogy Section, Copenhagen, Denmark. As in the previous paper, the researchers used Microsoft Kinect with multiple sensors to scan the face completely and precisely. In this paper, a multimodal natural user interface system, based on real-time audio, video and depth processing, was demonstrated. To illustrate the concept, they have four possible scenarios: 
\begin{enumerate}
\item Login via face detection system, which we have seen recently in Windows 10; 
\item Application selection via object detection-recognition; 
\item Authorization control according to log in and data, and; 
\item Application operations. 
\end{enumerate}

The input devices consist of an RGB camera, and depth and audio sensors, with each device working independently without hampering the rest of the system. The system architecture implements a multimodal system based on natural user interaction as shown in Figure \ref{fig:SimpleUserRights}. The multimodal process involves face detection, objection detection, speech recognition, and gesture recognition. The face detection is completed in two steps. first, the joints of the head are tracked by the Microsoft input received. This cropped image does not provide sufficient data for user recognition since joints are not always stable which leads to the use of a two-way authentication process. A face detection algorithm is used to extract recognition data from the cropped image as shown in the Desktop Login Flow Diagram in \ref{fig:DesktopLoginFlow}. This combined method is then used in the future, that is, authenticating the user when they attempt to login on subsequent occasions \cite{vidakis2012multimodal}.

\begin{figure}[h!]
            \centering
            \includegraphics[width=0.5\textwidth]{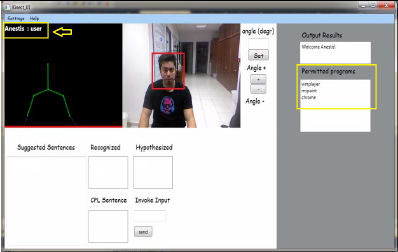}
            \caption{Simple User Rights Login}
            \label{fig:SimpleUserRights}
\end{figure}

\begin{figure}[h!]
            \centering
            \includegraphics[width=0.5\textwidth]{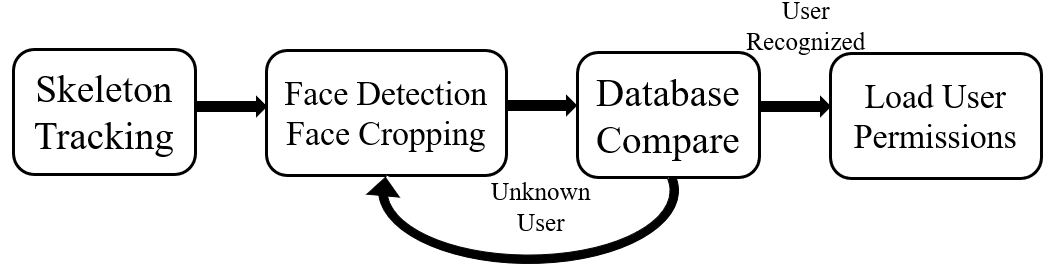}
            \caption{Desktop Login Flow Diagram}
            \label{fig:DesktopLoginFlow}
\end{figure}

There was no empirical data provided, but in all four phases, experiments were performed, and the user was able to log in, use an application, and perform operations using Google Chrome and MS Paint [44]. Another niche concept is explored, where intention recognition is used, a fairly novel idea, which includes intention recognition in conjunction with multimodal systems in the year 2015 via research at the Electronics and Telecommunications Research Institute in Daejeon, South Korea,  resulted in a paper entitled "Multimodal Data Fusion and Intention Recognition for Horse Riding Simulators." The research developed a system that gives a user the feeling of riding a horse and attempts to teach the user the skill of horseback riding. As it is not possible or feasible for everyone to learn horseback riding on an actual horse, this prototype enables users to learn and ride within a simulated environment. The proposed system consists of multiple data acquisition components, a feature extraction component, and a data fusion component. The system can increase realism for the user by enabling riders to perform interactions similar to ones they would perform while riding a horse.

The hardware consists of a multimodal user interface, multiple cameras, microphones, and other sensors to capture the user’s natural speech and movements. Three kinds of sensors were used namely contact, auditory, and two visual sensors. One contact sensor was mounted on the body of the simulator which senses that the user is riding the horse, while the user wore additional contact sensors. The auditory sensor was mounted on the helmet of the horse-riding simulator and captured the voice commands from the user. Two depth-sensing devices were used for capturing visual information. The contact information captured by the contact sensors included balanced sitting, drawing or pulling reins, spur, whip, and so forth. Using "Gesture and Speech Control for Commanding a Robot Assistant," the researchers performed experiments used a robot called ALBERT \cite{rogalla2002using}. They conducted experiments using gestures by considering the heuristics of hands by utilizing a webcam available on the robot. The experimental results show 95 percent correct recognition of hand gestures displaying yes and no (thumbs up is considered a yes while thumbs down is a no). Verbal Input “ViaVoice”, which is a speech recognition software offered by IBM, is used. This research does not deal with fusion at all; the experiments performed separately.

\begin{table}[h!]
\centering
\caption{Gesture Phrase Distributions Per Recording}
\label{table:7}
\begin{tabular}{|c c c c c c c c c c |} 
 \hline
   Red. Id	& g1	& g2	& g3	& g4	&g5 & 	g6	& g7	& Total & Dur (s)\\
\hline
1	& 52	& 64	& 9	& 22	& 1	& 0	& 19	& 167	& 239\\
\hline
2	& 20	& 40	& 1	& 8	& 0 &	17	& 6	& 92	& 167\\
\hline
3	& 22	& 49	& 1	& 23	& 10	& 21	& 40	& 166	& 265\\
\hline
4	& 53	 & 60	& 15	& 20	& 4	& 18 & 	20	& 190	& 347\\
\hline
5	 & 2	& 45	& 1	& 0	& 0 & 	0	& 0	& 48	& 155\\
\hline
Total & 149 & 	258 & 	27 & 	73	& 15	& 56	& 85	& 663	& 1173\\
\hline
\end{tabular}
\end{table}

In another experiment, intention recognition is the niche concept explored while using horse-riding simulators. This idea is completely new and has not been discussed elsewhere in the literature surveyed herein. Intention recognition is implemented by defining a class that stores every input received and updates the database whenever it receives a new input. For every action, an intention class is defined. For example, the balancing intention class captures the actual position of the user, and the exercise maintenance intention class corresponds to the leg release or bridle release actions. The main intention class includes strength information expressed through the action as show in Figure \ref{fig:SampleCues }. Once all the data is collected, it is compared with data from the intention database. Based on the results, the system recognizes the intention of the user. The researchers have included multimodal data fusion but have not shared any empirical data about the accuracy of the system \cite{kang2015multimodal}. 

Building a multimodal human-robot interface, a paper published in 2001, talks about how to build a system which can accept multimodal inputs. The author talks about personal digital assistants as a form of input apart from speech and gesture \cite{perzanowski2001building}. The media center application designed, while considering the requirement of the differently abled users, and it exploits minimal hardware, namely. PC (Athlon X2 3800) running the media focus server programming, a cell phone (Nokia N95) for interfacing with the client and running the customer programming, a remote get to indicate associate these together, and a superior quality forty-inch advanced TV showing the UI.

\begin{figure}[h!]
            \centering
            \includegraphics[width=0.5\textwidth]{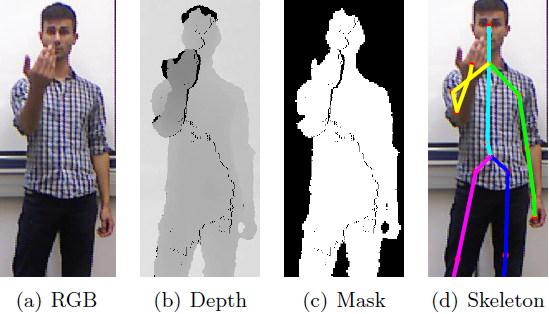}
            \caption{Sample Cues of the Multimodal Gesture Challenge 2013 Data Set}
            \label{fig:SampleCues }
\end{figure}

Skubic et al. (2004) designed a system called Spatial Language for Human-Robot Dialogs which enables the robot to analyze surroundings and managing things around it. In an example, it has been illustrated that the robot can analyze things around it and respond verbally where things are located. Spatial language has been used to define the geographical locations of objects lying in a room. The idea has been discussed though it has not been implemented and no statistical data is provided in the paper. An example is illustrated in the dialogue below \cite{skubic2004spatial}.
\begin{enumerate}
\item Human: “How many objects do you see?” 
\item Robot: “I am sensing seven objects.” 
\item Human: “Where is the box?” 
\item Robot: “The box is behind-right of me. The object is close.” 
\end{enumerate}

Multimodal Media Center Interface based on speech, gestures, and haptic feedback was designed in the year 2009 in Finland. The proposed solution consists of a multimodal media center interface based on speech input and haptic feedback. The system architecture contains a zoom-enabled, context-focused GUI, tightly coupled with speech input. The core idea of this research is to make a substitute system for trivial digital home appliances, such as, remote controls, game controllers, and so on. The author has argued that many of these interaction devices are not user-friendly. A classification is made based on the visual ability of the user. For a blind user, speech and haptic inputs are sufficient for accessing the information, while a zoom-enabled GUI is proposed for visually impaired (low-vision) users \cite{turunen2009multimodal}. The application was developed using C Sharp and ran under Windows XP. The application consists of speech recognition while the mobile device can interpret the gesture, speech, and haptic feedback through a GUI. The mobile technology was based on a native Symbian application while the GUI and main logic used MIDP 2.0. The media center provides an electronic program guide which enables the user to access complete digital television content. The system consists of two graphical displays: a television and a mobile phone display. The first GUI on the television uses a matrix format to display the interface, explain its usage, and presently available content. One other proposed solution is to mount a wireless microphone instead of using a mobile phone for physically challenged users. It is unknown when or if the proposed system will be implemented.
Assistive Robots for Blind Travelers is an ongoing project at the CMU Robotics Lab in which a robot is attempting to guide visually impaired people through an urban environment. For individuals with disabilities, transportation remains a major barrier for living a quality of life. With the advent of robots, it could be argued that their life would be much easier, especially with the usage of smart buses and shuttles. The differently abled people can live a healthy life, but they cannot drive, eventually makes their life tough. The visually-impaired can use the systems based on physical, verbal and digital input defining the foundation of human-robot interaction. The objective of this project is to enhance the safety and independence of visually impaired travelers. 

The implementation involves the following three pieces \cite{Buffomante2017}: 
\begin{enumerate}
\item Rathu Baxter: Rathu Baxter was originally designed to assist human manufacturing settings 
\item Mobile Robots: Mobile Robots can enhance the navigation experience of blind and visually impaired travelers in urban environments.
\item NavPal: NavPal is a smartphone app to give navigational assistance to blind adults as they move around unfamiliar indoor and outdoor environments. As it is early in the implementation phase, not many details are available. 
\end{enumerate}

Human-speech perception is a multimodal process which provides higher knowledge resources such as grammar, semantics, and pragmatics. The information source, which is used in the presence of noise, is lip-reading or also known as speech-reading. Automatic speech recognition (ASR) is a very active research area for several decades, but despite the fact many teams are working on it still would not able to compete with the performance achieved by human ears: the results achieved by ASR systems are far lacking from expected results. Most state-of-the-art ASR systems use acoustic signal only and ignore visual speech cues. Therefore, they are susceptible to acoustic noise, and all real-world applications are prone to error because of some noise in the background \cite{dupont2000audio}. 

Implementing the concept of multimodal human-robot interaction framework, a personal robot was designed at Universidad Carlos III de Madrid. The architecture, used in developing the prototype, is Automatic Deliberative, which incorporates an emotional control system (ECS). The Automatic Deliberative (AD) architecture is based on a human psychological model. In the deliberative piece, the robot can do tasks such as planning and word model management, while the automatic piece pertains to reactive and sensory skills. An emotional control system is added to this Automatic Deliberated Architecture, as shown in Figure \ref{fig:Automatic_Deliberated_Architecture}. By using this architecture, the researchers would be able to train the robot in skills such as greeting, face recognition, user identification, dialogue, audiovisual interaction, non-verbal visual expression, and dancing. The system designed has been named Maggie. There are three modalities that the researchers have proposed for use: visual, voice, and audio-visual mode.

\begin{figure}[h!]
            \centering
            \includegraphics[width=0.5\textwidth]{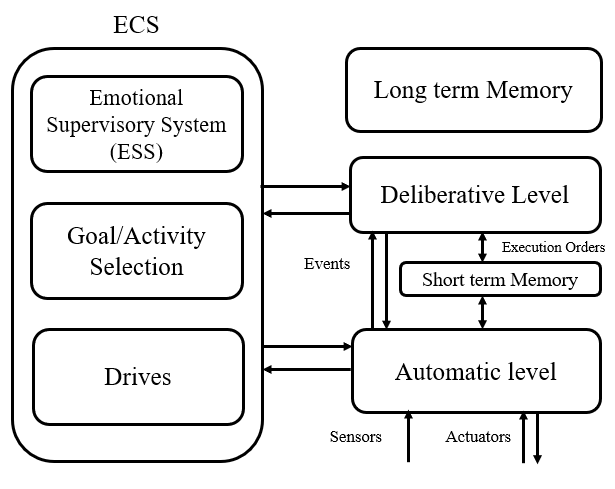}
            \caption{Automatic-Deliberated Architecture}
            \label{fig:Automatic_Deliberated_Architecture}
\end{figure}

Visual mode is enabled via the Proxemics and Kinesics expression control; the voice mode uses the Text to Speech library VTxtAuto \cite{gorostiza2006multimodal} (VoiceText 1.0 Type Library) to generate speech from text. For the audio-visual mode, images and sound expression input is provided through a tablet PC, and Maggie understands it by utilizing Pure Data and Graphic Environment for Multimedia (PD-GEM), which is an open source audiovisual software tool. In this research article, the authors have not presented any empirical data that quantifies the accuracy of their robot. As per our understanding, it is just a hypothesis that explores the possibility of using multimodality in robotics. The proposed framework could perhaps be used to implement game scenarios and choreography programming.

Multimodal input from a robotic arm is implemented to perform grasping, unscrewing, and insertion tasks on a Barrett's robotic arm. The inputs involved are multimodal sensory signals, and it achieves  80-90 percent while performing the tasks. The drawback of this system is that it only worked on those trajectories that have been shown to the robot earlier, which implies it works in a constrained environment \cite{su2018learning}.

So far, we have seen various applications of multimodal data fusion, including automation in health care domain. Another interesting experiment was performed in Germany in the year 2018. A device is used with sensors to capture multimodal data for the detection of user’s motion intention and its assimilation into the exoskeleton control system while climbing the stairs or walking. The systems claim to achieve the average accuracy of 92.8 percent using the Hidden Markov Models (HMMs). While analyzing the system it’s been observed the placement of IMUs, and force sensors are used for capturing the data. Moreover, the details of fusion are not explained explicitly. The algorithm used is unable to predict a deeper analysis of the latencies for different motion transitions \cite{beil2018human}.

The application of multimodal systems varies from assigning tasks in an industrial environment to help humans in climbing the stairs. Along similar lines, a concept was introduced called mixed reality, in which Microsoft HoloLens is used to perform Pick-and-Place tasks on things placed on the ground. The experiment has achieved an accuracy of 93.92 percent, but it will only perform simple trajectories. The experiment demonstrated Microsoft HoloLens camera is mapped with the motors of the robotic arm and perform actions. The idea is not very new, but the implementation needs to achieve the accuracy for complex trajectories \cite{krupke2018comparison}.

\begin{table}[h!]
\centering
\caption{Human Sensory Modalities Relevant to Multimodal Fusion}
\label{table:8}
\begin{tabular}{|c c |} 
 \hline
  \textbf{Modality} & \textbf{Example} \\
 \hline
Visual	& Face Location\\
\hline
 & Gaze\\
\hline
& Facial Expression\\
\hline
 & 	Lip Reading\\
\hline
 & Face-based identity (such as age, sex, race etc.)\\
\hline
& Gesture (head/face, hands, body)\\
\hline
& Sign Language \\
\hline
Auditory & Speech-input\\
\hline
& 	Non-Speech audio\\
\hline
 Touch & Pressure\\
\hline
& Location and selection\\
\hline
& Gesture\\
\hline
\end{tabular}
\end{table}

\subsubsection{Computer}
One device that has continued to be upgraded and improved through the years since it was first created has been the computer. One can argue when it was first created which has led to the discovery of two possible starting points back in 1622 when William Oughtred created the first blueprints for a very crude-looking computer. One the other hand, in 1833 and 1871 when Charles Baggage created the first computer that resembled our modern ones, and it was called the Analytical Engine. With the computers going as far as the 1600s, it is insane to see how advanced and powerful it has become when back then one could only do the simplest of functions or commands with it. In current times, these have become one our staple points as so many people use them these days, not just for the latest tech but also for the internet and what we can do with it. Also, how computers were mainly used for research or creating new software and not for the personal joy of others till many years later now where almost an average American family will have some form of a computer in the household. Then there is the issue of how teachers are using them to help teach younger students, as since the internet has more to learn whereas it also has flashy things that can get young ones attention while learning new things. As to how far we have come with technology, it also comes with new threats in the digital zone as people can hack into bank account, and steal a whole person’s life away with a single click. The fear of being hacked virus making its way into a computer or device leads their creators to try to develop the latest software defenses to protect them. While some can hold hackers and viruses at bay for some time, the other side is also trying to improve their methods to ruin people's devices or lives knowing how addicted modern people are to them \cite{WhoInventedComputer:online}. PCs accepts input from various peripheral devices such as a mouse, keyboard, microphone, touchpad, webcam, and fingerprint scanner. All of them work independent of each other, which leads to conclude that a personal computer is a multimodal device without fusion.

\subsubsection{Cellphone}
A very common device in our modern times that has upgraded since the very simple phones from the 1900s is the cell phone. The very first cell phone was created for the world in 1983 when the company Motorola launched the DynaTac 800x for only \$4,000 one could get a portable phone with a battery lifespan of 30 minutes, but back then this was revolutionary as it enabled people to walk and talk without being limited by a wire or cable. In modern times, the common cell phone has gone through so many phases by different companies that it is impossible to list them all, but nowadays they are mainly known by the best shape to hold in your hand or pocket, the best camera, the fastest data, and the newest and most trendy phone in that year. Some people would say that cell phones are in pretty much everyone's hands, and the companies do not show any sign of slowing down in their race to create the best phones and make the most money. The older phone are less fragile, whereas new phones are prone to corruption or are very fragile. That is one of the main reasons many Android users do not want to get an iPhone they can break very easily, and many are just slightly upgraded copies of their past versions. Many iPhone users love the smooth feeling and fancy covers they can buy for their phones, but the majority do end of chipping or shattering the screen, which are expensive to fix \cite{Cellphone:online}. Sadly, many people are easily drawn into the buying circle of these phones as commercials usually depict some person with the company’s latest phone having a good life or being the center of attention while taking pictures. Too many people, especially teens, are easily drawn into buying them to follow the trend, and this will continue until they either run out of money or the company stops producing phones. Some companies, though, will ask people to take surveys on what they want from their phones and try to meet demand so they can profit off them. Surveys indicate that what people want most from phones: a better look, faster access to internet, a nicer feel, and more storage. So companies will try to meet demands, and some do a pretty good job of upgrading their devices: others will keep the same format and make one minor upgrade and then sell the phones for a higher price. The latest cell phones in 2019 are capable of accepting a user's input from touch and speech. The idea of implementing error-prone device is still far away; on the contrary, most cell phone is keep track of human activity throughout the day, which is the biggest danger to the freedom of the user. The google maps application tracks all the location visits, and even if the user disables them, Google will keep the data for a month before deleting it. The health app captures the number of steps walked in a day and how many stairs a person climbs. Moreover, it makes it more susceptible to cyber attacks, not only in terms of money but also loss of information. Companies advertise their products based on lifestyle, for example, a user may have an issue with blood pressure; the person's cell phone will capture his or her details, and pharmaceutical companies will  start offering insurance and medicines. 

\subsubsection{Apple Smart Watch}
The advent of smart watches has brought revolution into our lives; millennials and teenagers are comfortable using the latest technologies. The Apple smart watch accepts touch-based input from the user. The new Apple smart watch \ref{fig:AppleSmartWatch} has many different features. These watches have features such as GPS, a heart sensor, and a speaker.  The smart watch can access applications on your iPhone, such as messages and the camera. These watches are linked to your iPhone, which makes it convenient to answer phone calls and messages. Apple watches start at \$399. Apple watches can be put on many different kinds of bands made out of leather, metal, or nylon. The new Series 4 watches are a little bit larger and a little thinner than the previous models. The new Apple watch has up to eighteen hours of battery life and can be wirelessly charged. These watches are great for notifications and phone calls. The new version also has Bluetooth  built in and can pair with Bluetooth headphones and play music straight from the watch. Emails, phone calls, and text messages are easy to respond using this watch. Voice commands are also available, which make sending messages easy. Alarms can also be set, and the watch rings and vibrates just like a phone. This watch does it all, but it is a bit pricey for a watch. Siri responds to most common questions on the watch as well but does take some time to receive info from the phone. The Series 4 watches can come with either GPS or GPS and cellular data \cite{AppleSmartWatch:online}.  

 \begin{figure}[h!]
            \centering
            \includegraphics[width=0.4\textwidth]{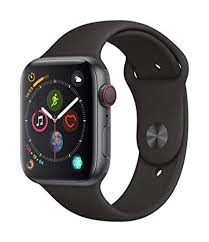}
            \caption{Apple Smart Watch}
            \label{fig:AppleSmartWatch}
\end{figure}
	
\subsubsection{Microsoft Modern Keyboard with Fingerprint ID}
Since we first began using PCs, one of the main ways we could interact with computer and input information through a QWERTY keyboard.  We still use QWERTY keyboards just as frequently, although there have been some slight changes to them over time. There are ergonomic keyboards for comfort and prevention of carpal tunnel for those whose spend a lot of their day typing. Some keyboards connect wireless to your PC, eliminating some of the wire clutter. Back lit keyboards make it easier to see, especially if you are typing in the evening or where there is low light. One can get a keyboard that has a track pad on it, eliminating the need for a mouse. Many of the keyboards, especially those made by Apple, are very thin. Many keyboards now have customize shortcut keys. With the same retail price between Apple and Microsoft \$129. However, the Microsoft key broad use the newest technology which is finger print sensor. The finger sensor use asymmetric key cryptography, which creates more secure environment for user when they use it. Reduce cumbersome password. The key board connect to our computer by Bluetooth function.
This Microsoft keyboard as shown i Figure \ref{fig:MicrosoftModernKeyboard} stood out more than the others because it has Bluetooth and a USB connection for recharging the battery. It was designed for comfort; it also has an added feature that is not often seen. It has biometric security included with a hidden fingerprint scanner for an extra secure but easy login option if you are running a Windows 10 operating system on your PC \cite{MicrosoftKeyboard:online}.

\begin{figure}[h!]
            \centering
            \includegraphics[width=0.4\textwidth]{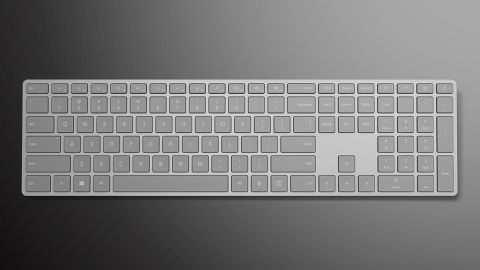}
            \caption{Microsoft Modern Keyboard}
            \label{fig:MicrosoftModernKeyboard}
\end{figure}

\subsubsection{Autonomous Vehicles}
Most of the smart vehicles of this era have capabilities such as communication with other vehicles, communication with the infrastructure (traffic signals and traffic update), GPS, sensor-driven decision making, etc. Connected autonomous vehicles (also known as smart cars) are driverless, capable of making their own decisions based on data from various sources (with little or no input from the user) and avoid obstacles that come in its way without causing any discomfort to the passengers or the other cars near it. These cars gather data from a myriad of sensors, the internet, roadside infrastructure, GPS, and so forth, and it is fed to the driving model which makes crucial driving decisions. Communication plays a significant role here, as most of the data coming to the smart car are from other smart entities, such as other smart cars and roadside infrastructure.

The self-driving car has both intrigued and terrified people over the years because of technical glitches observed and accidents that have happened on the road. If we move to the history of autonomous vehicles, one the very first autonomous cars was invented in 1925 when inventor Francis Houdina created a radio-controlled car that could start, shift gears, and perform most of the functions available that time while Houdina never touched the wheel. From 1925 to 2019, the progress of self-driving cars has continued to improve, and many hope they will become a reality without fear of the AI taking over and running over people. Sadly, in 2014, the first self-driving car fatality occurred when a Tesla tester died when the car hit an eighteen-wheeler. This has sparked a huge debate on whether people should continue to fund these projects or just let them fade away. However, many argue that if given enough time, the dream of autonomous cars can become a reality and may prove far safer than regular cars. One argument was how it could help people who were intoxicated and unable to drive can simply get in, and the vehicle will take them home without fear of crashing. Many creators and developers of self-driving cars want to keep pushing people to trust them and help them bring these futuristic vehicles to life. Surprisingly enough at CES 2018, where they announced the latest cars to be released or teased for the upcoming years. One company by the of name Nvidia revealed an autonomous car named Xavier where they shall incorporate AI into it. Many people who attended the event were very excited to see the self-driving car quest still going and now looks very promising with our current technology, many hope for the satisfaction of relaxing while the car drives them wherever they need to go. The company has also put out a teaser on how users can program the car either by typing in their destination or speaking the location to the google maps or any other navigation system. \cite{SelfDriving:online}. The first autonomous vehicle we are going to be discuss is Tesla.

\paragraph{\textbf{Tesla}}

Tesla is a self-driving car made by Tesla and Elon Musk. Tesla was founded in 2003. This brand was created with the mission of the new technology. Using sensor and electricity to run the car. No CO2 release to environment, modern, convenient. Tesla was originally founded in 2003 by a group of engineers. Tesla was made not only for revenue purposes but also to motivate and influence the consumers to use zero-emission cars, which is a topic of great interest these days. With global warming and rising gas prices, more people want to transition to electric cars, but Tesla as shown in Figure \ref{fig:Tesla 3D layer} is not only an electric car but also self-driving. With this type of technology, HCI is inevitable. Just because the car could drive by itself does not mean a human is not important during the interaction process. A human is needed to tell the car where needs the human to pay attention at all times to avoid accidents from happening. Self-driving cars such as Tesla have multiple sensors place all around the car that help the car understand the environment so it can steer itself appropriately Model the 3D motion based on reality model which optimize segmentation. Analysis those segmentation into extract 3D motion, that contains a lot of layers and map out geometrically those motion \cite{menze2015object}. The car will output these code and data set to self-drive. The car has a high-precision, digitally controlled electric braking system, twelve long-range ultrasonic sensors, a forward-looking camera, and forward radar. The ultrasonic sensors are placed around the car so they can sense sixteen feet around the car. They sense when something is too close and also used for lane changes. So far, the interaction is using the touch screen-based tablet embedded on the dashboard of a smart car. Despite the evolution of technology, we have not reached a level where a customer can fully rely on a car; he or she needs to be attentive at all times while using it \cite{Tesla:online}.  The user has the freedom to operate the car using the traditional way, or it can set on auto-pilot temporarily.

\paragraph{\textbf{Pal-V}}

Pal- V stand for Personal Air and Land Vehicle is a Dutch company that is one of the first company produce flying car. PAL-V car is the combination between a car and a auto-giro. The driver has to have the license of driving car and driving auto-giro license.This car was produce for people who want to own their own plane but not as an enormous size.The Pal-V Liberty is another autonomous vehicle which used the identify technology- 3D layer motion, but what makes this one so special is that it can fly as well. The interaction with the user is the most important part, which we will discuss apart from other features. The vehicle is unique because it has two separate engines for flight and another for driving. This vehicle is capable of going to a maximum speed of 100 mph and takes ten minutes to transform into the driving mode or flying mode. When the Pal-V Liberty as shown in Figure \ref{fig:Pal-V_Liberty} is in flight mode, it could reach speeds of 112 mph with a maximum range of 817 miles. When the cae is in drive mode, it is 4 meters long, 2 meters wide, and 1.7 meters high. In flight mode, the vehicle measures 6.1 meters long, 2 meters wide, and 3.2 meters high.  These vehicles cost between \$399,000 to \$599,000 and will require a pilot's license to own or drive. Because someone would need a pilot's license to fly, it makes it more complicated not only as a product but also legally. The responsibility and complexity of not only driving but having to fly a vehicle such as this one requires a human to have full focus and attention to operate it, especially when in flight mode. The Pal-V Liberty is made of hand-laid carbon fiber parts, cockpit leather, lightweight aviation aluminum, and an electrical system. It also runs on premium e10 gasoline and gets thity-one miles per gallon in car mode and 6-9 miles per gallon while in the air. The Pal-V liberty was in the works since 2008, with a successful prototype completed in 2009 and the second prototype developed in 2010. It was shown at the Geneva Motor show in March 2018. A simple flying automobile need the presence of a certified driver as well as an aircraft pilot at the controls. For the vast majority of humans, this is unfeasible, hence computer technologies that make flying easier will be required. These operations include aircraft movement, navigation, and emergency procedures, all of which are performed in potentially congested airspace. Fly-by-wire computers can also compensate for problems with flight dynamics, such as stability. A feasible flying automobile could have to be a completely autonomous vehicle with just passengers. \cite{PAL-V:online}.

\begin{figure}[h!]
            \centering
            \includegraphics[width=0.4\textwidth]{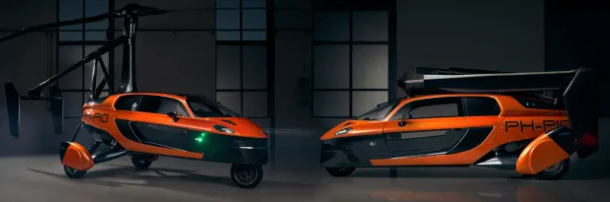}
            \caption{Pal-V Liberty}
            \label{fig:Pal-V_Liberty}
\end{figure}

\section{Discussion}

Nowadays, technology is develop substantial and massive, the use of technology is an essential part of our lives, however, the daily use of devices brings challenges to technology. The more human involving  to technology, the more abundance and challenge the high-tech network the scientists had to clarify. There are many instances which sometimes the systems capable of handling speech input fail at the stage the user needs technology the most. In such scenarios, the use of highly sophisticated gadgets become a nightmare for the user. There are two broad categories in which users are classified: first, people who are well versed in technology, and second, the senior citizens who are not well versed in technology. The second category finds themselves in huge trouble if they have to interact with technology, and they have to rely on it completely. The scientist and inventor come up with the simpler ideal will is AI technology which is the use of voice assistants. It has increased significantly during the last few years, such as Amazon Alexa, Apple’s Siri, Google Assistant, and Cortana from Microsoft. These assistance recognize it owner by voice tone and give us the information based on own demand. However, sometimes, those assistance could not input the right data since they could not recognize human voice well enough. The big IT companies are trying their best to compromise the problem recognizing the voice commands inaccurately. Moreover, other modalities such as touch and gesture come with their own set of challenges and a cognitive load to learn the systems. The need of this hour is to design a  system and capable of managing the input from various users. Here area few challenges that need to be considered while designing the multi modal system.

\subsection{Challenges while designing Multimodal Systems}
Designing a multi modal systems is always challenging. The designer should have a broad picture of what kind of requirement they are going to address and who is the user base. Oviatt’s “\textit{Ten Myths of Multi modal Interaction}” published in 1999 provide useful insights for those planning to develop a multi modal system. 
\begin{enumerate}
\item A achieved multi modal system is the system that the users will interact well with it. Well-designed systems are set up to allow users to choose their preferred modality for system interaction, which may be uni- modal or multi modal.
\item Speech and gesture are the dominant multi modal integration pattern.
\item Multi modal input involves simultaneous signals. Multi modal systems should be designed to allow for sequential use of modalities rather than simultaneous use. 
\item Multi modal integration involves redundancy of content between modes. Usage of varying inputs is preferred over using a single modality again and again. 
\item Enhanced efficiency is the main advantage of multi modal systems. However, multi modality does not necessarily increase efficiency; it may or may not. They are designed to provide increased flexibility and increased user satisfaction. 
\item Individual error-prone recognition technologies combine multi modalities to produce even greater unreliability.
\item All users’ multi modal commands are integrated uniformly in the same technology system, such as: Apple system or Window system.
\item Different input modes can transmit comparable content.
\item Enhanced efficiency is the main advantage of multi modal systems.
\item There are multiple input method, however, speed is the mandatory. Most of multi modal system used speed as an input.
\end{enumerate}

\subsection{Proposed Multimodal System Design guidelines}
Reeves et al. (2004) proposed the guidelines for multi modal systems. Here are the mentioned proposed guidelines \cite{reeves2004guidelines}:
\begin{enumerate}
\item Multi modal systems should be designed while keeping in mind the broadest environment a person could encounter while using the system, for example, use in a private office vs. while driving a car.
\item The designer should consider privacy issues while accessing the system. For example, perhaps a person should be prevented from using speech/voice input while using the system publicly as doing so could lead to a breach of personal information. 
\item Maximize human cognitive and physical abilities. The multimodal interface should be designed in such a manner as to be easily understood by the user. 
\item Modalities should be integrated in a manner compatible with user preference. For example, users have provided input via speech, they should have the option to receive the output via whatever modality they wish and not just speech. The system should be able to be customize according to user needs. 
\item The multimodal system should adapt to the needs and abilities of the user. Individual differences such as age, preferences, pronunciation, sensory skills, and so forth should all be accounted for while designing the system. 
\item The output should be consistent and prompt. 
\item The system should provide a robust error-handling mechanism. 
\end{enumerate}

Later in the research, the author discussed multimodal integration and whether it should be performed early or late in the development process. There is no clear-cut answer given to this issue. It varies from system to system. But the authors also presented a ‘big picture’ which explains what the output of multimodal integration should look like. The system should be able to manage discrete events, loops, and handlers timely and in ways that  better match the human interaction the system is intended to support.

\subsection{Table of Multi Modal}
Technology can emphasize the connection between human and machine. Through this paper, we demonstrated and explained the variety of techniques available in the market. We have mentioned many modality concepts, such as gesture, speech, face ID, voice, haptic. Through the paper, those modality concepts have been recorded and explained clearly. Table IX below will present the comparison of multi modal technology. It will summary the modality was used in each experiment and how accurate it will be. Depend on the experiments, each of them will have different features and background information. The table also mentioned the technology which has been approached in each project. Each experiment has different multi-modal inputs and different technology. Each experiment has different multi-modal inputs and different technology. For instance, column 3 and column 4 in table IX present the variable of multi-modal. The table summary all the experiments were mentioned and analyzed before; it makes the reader can have a better concrete idea of the paper.

\clearpage
\onecolumn 
\begin{landscape}
\begin{center}
\LTcapwidth=\linewidth
\begin{longtable}{p{30pt} p{50pt} p{50pt} p{50pt}p{90pt}p{40pt}p{50pt}p{70pt}p{70pt}}
\caption{Comparison of Multimodal Technologies}
\label{table:typology}
\endfirsthead
\endhead

\textbf{Author} & 	\textbf{Experiment} & 	\textbf{Modality Used} &	\textbf{Multimodal Input} &	\textbf{Technology Used	} & \textbf{Fusion}	& \textbf{Accuracy Claimed}	& \textbf{Features} & \textbf{Drawback}\\ 
\hline
\endhead

\textbf{Author} & 	\textbf{Experiment} & 	\textbf{Modality Used} &	\textbf{Multimodal Input} &	\textbf{Technology Used	} & \textbf{Fusion}	& \textbf{Accuracy Claimed}	& \textbf{Features} & \textbf{Drawback}\\ 
\hline

 Michela & 	ECOMODE \cite{ferron2015mobile}  & 	Haptic 	& X	& Samsung Galaxy S5, iPad mini, and a Samsung Galaxy Tab S 10.5 & 	X	& No Empirical Data & 	Touch screen devices used for Elderly people & 	The experiments do not yield results in low lighting conditions.\\ 
\hline

Ian &	Baxter Research Robot\cite{lenz2015deep}&	Gesture & X & Microsoft Kinect, Baxter robot, PR2 (“Kodiak”) & X & 84 - 89\% & X & 	Detect things which are in front of the robot only.\\ \hline

Natalia	 & ModDrop\cite{neverova2015moddrop}  &	Gesture, Speech  & 	Y 	& Audio and Video files & 	Y	& 96.77\%	& Multimodal deep learning used for fusing two inputs voice and gesture. &	X\\ 
\hline

Harold &	Donaxi\cite{vasquez2015using} & 	Gesture & 	X	& The robot, and Microsoft Kinect 2 & X	 &	~2000 iteration required for training  &	Omni directional navigation system &  	Extensive training is required before using the system.\\ 
\hline

Finale &	MIT's Human-Robot Interaction\cite{doshi2008spoken} & 	Speech & 	X &	Speech Recognizer (ASR), Natural language (NL) parser, and a dialogue manager (DM)	& X & 	76\%	& Human speech implemented on a wheelchair 	& Only accepts speech \\
\hline

Mathieu	& CommRob\cite{vallee2009improving} 	& Speech, Gesture & 	Y & 	Java-based FreeTTS API, CommRob Robot	& Late Fusion & 	100\% (under restricted conditions)	& Speech and Gesture are combined to derive moving of robot at (x,y) coordinates	& The experiment did not mention explicitly how they have achieved fusion. \\
\hline
Cristian	& Modeling of interaction design\cite{bogdan2008modeling}	& Gesture &	X & 	UI based Modeling tool & 	X &	A satisfactory response in the 4th iteration & X & X \\	 	 
\hline

Haleh &	Combining Voice and Gesture \cite{chizari2013combining}	 & Speech, Gesture	& Y &	Microsoft Kinect API& 	Y	& 99.17\% &	Multimodal Systems proves to be better than Unimodal Systems & 	Hard to recognize the sound, not reliable and the expeiment can be tiring while using both hands and speech\\
\hline

Nikolas &	Natural user interaction\cite{vidakis2013multimodal} & Speech, Gesture &	Y &	Microsoft Kinect with RGB, depth and audio signal.  & Y &	No Empirical Data &	Proposed architecture could be used with any application running on a computer. Used for moving slides in PowerPoint  &	Does not have the ability to understand the usage of grammar to understand the input.\\
\hline

Emanuel	& MaRDi Dialogue system\cite{ferreira2015reinforcement} &	Speech, Gesture & Y & 	3-D simulation software MORSE, KTD-Q algorithms & Y &	93.73\%	 & The Pick-Place-Carry task in a Human-Robot Interaction &  	Interactions take place in a simulated 3-D environment where the user appraisal acquisition is simplified \\
\hline

Emanuel &	TownInfo\cite{ferreira2015reinforcement} & 	Speech, Gesture &	Y &	Simulator using KTD-Q algorithm &	Y & 	95\%	& A virtual tour guide, better robustness to noisy conditions in terms of semantic input error rate & 	\\
\hline

Angelica	& Motherese\cite{lim2014mei}	& Speech, Gesture, and input file &	Y &	Multimodal Emotional Intelligence (MEI) with SIRE input (Speed, Intensity, irRegularity, and Extent). & 	X	& 72\%	& The Motherese robot works with these types of input which are .wav, Vocaloid, Kinect, NAO, Keepon, Flute, Theremin &	Accuracy will be improved by training the system in a more diversified environment, e.g., a 20-dimensional confusion matrix.\\
\hline

Tetsushi &	Manipulator Control Interface \cite{oka2015multimodal} &	Speech, Gesture &	Y &	7-DOF manipulator (iARM), laptop with a Touch Screen, a CAN-to-USB adaptor, a USB headset & X &	No Empirical Data &	The robot can easily pick up, rotate and replace objects using gesture and multimodal commands &	A user would not be able to operate effectively in a rotational mode\\
\hline

David & 	Navigation interface\cite{krum2002speech} & 	Speech, Gesture &	Y &	Virtual Geographic Information System (VGIS) and 3-D visualization environment &	X &	10.1 minutes for 100 commands &	3D navigation of the globe.  &	Users who do not have a clear accent will prefer to use gesture and pen.\\

\hline
Elif &	Arm Motion for Prosody-Driven Synthesis\cite{bozkurt2016multimodal}	& Speech, Gesture &	Y	& Hidden Markov Model & 	X &	67.80\% &	Used subjective evaluation methods
to set the system parameters and to assess animation quality over two different datasets. &	Doesn’t include semantic analysis of speech, synthesis of head motion and lip-sync, which would help to achieve more realistic animation results.\\

\hline
Julie	& Gesture and voice prototyping\cite{rico2010gesture}	& Speech, Gesture & 	Y & Social acceptability and User perception & 	X &	No Empirical Data	& Performance of audible and visible interactions, including how the user perceived the interaction and how comfortable they were while using the device. &	Results are not provided in Empirical form. A figure is shown with dots showing the understanding of input from various users. \\

\hline

Nikolas &	Multimodal desktop interaction\cite{vidakis2013multimodal}	& Speech, Gesture &	Y &	Microsoft Kinect, Microsoft Speech Recognition SDK.	 & X	& No Empirical Data & 	1. Login via face detection system, which we have seen recently in Windows 10
2) Application selection via object detection-recognition
3) Authorization control according to log in and data, and
4) Application operations	& The research idea is good, but results are collected in terms of successful attempts to open applications on a computer. \\

\hline

Rogalla & 	ALBERT\cite{rogalla2002using} & 	Speech, Gesture & 	Y &	ViaVoice &	X &	95\%  correct recognition &	The experiment uses heuristics of hands by utilizing a webcam available on the robot &	Research does not deal with fusion at all, both the experiments being performed separately\\

\hline

Sangseung &	Horse Riding Simulators\cite{kang2015multimodal}	& Speech, Gesture, Haptic & Y & Camera, microphones and other sensors. & 	X	& No Empirical Data	& Replicate the mechanistic movements of realistic riding motions &	Feature extraction, data fusion, and intention
classification is not explained explicitly. \\
\hline

Dennis	& Human-robot Interface \cite{perzanowski2001building} &	Speech, Gesture & Y & 	PC (Athlon X2 3800), Server programming, Nokia N95 and 40" advanced TV showing the UI	& X	& No Empirical Data & 	Works for blind users, visually disabled users, and physically impaired users. &	The robot requires specific input speech command to initiate the conversation. Moreover, the robot doesn’t work in noisy environments.\\

\hline

Marjorie	& Spatial Language for Human-Robot Dialogs\cite{skubic2004spatial} 	& Speech, Gesture &	Y &	Personal Digital Assistant, wireless microphone, interacts with a robot
via a touch screen and speech. 	& X	& No Empirical Data	& The robot will provide detailed spatial descriptions. &	Does not facilitate commands concerning objects in the environment, e.g., “Move forward until the pillar is behind you.”\\

\hline

Markku &	Interface Based on Speech, Gestures, and Haptic Feedback\cite{turunen2009multimodal} &	Speech, Gestures, Haptic  &	Y & 	PC, the Athlon X2 3800, Nokia N95, 40” high definition television, and a wireless connector. The application was developed using \# and ran under Windows XP. &	X &	Perceived quality of the speech
input surpassed the upper limit of user expectations	& Accepts speech input and the response was good. &	The interface would work for restricted input.\\
\hline

Stéphane &	Audio-Visual Speech Modeling \cite{dupont2000audio}&  	Audio-Visual, Speech &	Y	& M2VTS audio-visual database & 	X &	95.80\% & 	The database used is extensive with samples collected from 37 different users & 	The experiment deals with only Speech input. No other input from the user is accepted.\\
\hline

Chieh & 	Encoder-Camera-Ground Penetrating Radar Tri-Sensor Mapping\cite{chou2018encoder}	& Images and GPR/encoder data which are spatially evenly-spaced	& Y	& Camera, a GPR module which includes control unit, wheel encoder, and GPR antenna.	& Y &	98\% &	Developed a encoder-camera-GPR tri-sensor transportation infrastructure inspection sensing suite.	& Need to improve further speed and accuracy of the proposed algorithm.\\

\hline
Tom &	Road Marking Segmentation via Weakly-Supervised Annotations\cite{bruls2018mark}	 & Camera captures Images with annotated images.	& Y	& Camera captured images with annotated images in a weakly-supervised way. Experiment performed on Oxford RobotCar dataset. &	Y & 	75.04\%	& It has inexpensive manual labelling by exploiting sensor modalities. Useful in creating maps for autonomous vehicles.	&The experiment does not include semantic classification of the road markings to retrieve the rules of the road.\\

\hline
Shan	& Vision and Tactile Sensing for Cloth Texture Recognition\cite{luo2018vitac}	& Tactile images and vision & Y & 	GelSight sensor used for capturing camera images and tactile data. Deep Maximum Covariance Analysis (DMCA) algorithm is implemented &	Y & 90\%	& Calculated perception performance of either vision or tactile sensing. &	Temporal information is not included during the experiments. \\
\hline

Tian	& Prediction with Spiking Neural Networks for Human-Robot Collaboration\cite{luo2018vitac}	 & Gesture, EEG Signals, Speech	& Y &	MYO armband, Emotiv EPOC, Microsoft Kinect &	Y	& 88\% &	The experiment exploits unique implementation of Myo armband, Kinect and EPOC Emotiv devices in surgery.	& Does not include contextual information to improve early prediction capability e.g., the current status of task progress\\

\hline
Zhe &	Manipulation Graphs from Demonstrations Using Multimodal Sensory Signals\cite{su2018learning} &	Multimodal Sensory Signals &	Y	& Barrett arm and hand equipped with two BioTacs &	X	& 80 - 90\%	& Able to perform grasping, unscrewing, and insertion tasks on a Barrett's arm. &	The robotic arm didn’t work in those trajectories which are demonstrated earlier.\\
\hline

Jonas & 	Multi-Modal Sensor Data for Lower Limb Exoskeletons\cite{beil2018human}	& Multi-modal sensor data	& Y & 	Hidden Markov Models (HMMs) & 	X	& 92.80\% & 	Used for classification of motion patterns at each time step while climbing stairs. & 	Unable to conduct a deeper analysis of the latencies for different motion transitions.\\
\hline

Castro &	Tracking-based Depth Estimation of Metallic Pieces for Robotic Guidance \cite{di2018tracking} & 	Image, Lasers & 	Y & 	Kinect, Lasers, mono-camera	& Y & 	95\%	& Performs object recognition and tracking system in real time & 	The model developed is not used in any application; it just a prototype. To increase the usability, new utilities needs to be incorporated. \\

\hline
Dennis & 	Multimodal Heading and Pointing Gestures for Co-Located Mixed Reality Human-Robot Interaction	\cite{perzanowski2001building} & Speech and Gesture	& Y & 	Mixed reality interface implemented using Microsoft HoloLens & 	X	& 93.92\%	& The interface is capable of guiding a robotic arm to picks things &	Simple operation are performed, unable to investigate if more complex pick poses are requested. \\

\hline
Zhenzhou & 	Unsupervised Trajectory Segmentation and Promoting of Multi-Modal Surgical Demonstrations 
\cite{shao2018unsupervised} &	Video and Kinematic data	& Y	& Unsupervised deep learning network, stacking convolutional auto-encoder is used. & Y &	TSC-K is the biggest beneficiary with the improvement of 15.2\% on average	& The experiment was designed to accomplish tasks like needle passing and suturing during the surgery. 	& X\\
\hline

Luís &	Extended Bayesian User Model (BUM) for Capturing Cultural Attributes \cite{santos2018extended}	& Sensor simulator and Signal generator	& Y &	Experiment capture a unified representation of cultural attributes from heterogeneous information.	& Y & 	The framework has a significant impact on classifier precision over time, with an overall improvement of 27.88\%.	& Ability to learn user attributes from a distributed data stream, with increasing performance over time.	& Unable to find information on the user's current state, such as their mood, satisfaction level, etc.\\

\hline

\end{longtable}
\end{center} 
\end{landscape}
\twocolumn

\subsection{Future Idea for Human-Machine Interaction}

As the field is evolving day by day, the researchers are working hard to implement a robust and foolproof solution for daily use. In the era of the Internet of Things (IoT), most of the devices are connected with the network, including household devices such as refrigerators, fans, lights, and even our garages,which those simple things usually be controlled by remote device in Figure \ref{fig:Remote device media capabilities}. Following that model, the remote device connected to the Host PC 300 through the media capabilities. However, this is a elementary mechanise network since it shows the household devices transmission. We believe that User Interface will fundamentally change. The design of an operate system is not enable to accomplish yet because building high precision devices are expensive and the mechanise network are more complex\cite{shoemaker2008systems}. Moreover even the expensive components have probability of error percentage in every device. The technology scientists are tried to create more stability and accuracy with IoT based on the basic  diagram below. 

\begin{figure}[h!]
            \centering
            \includegraphics[width=4.5cm]{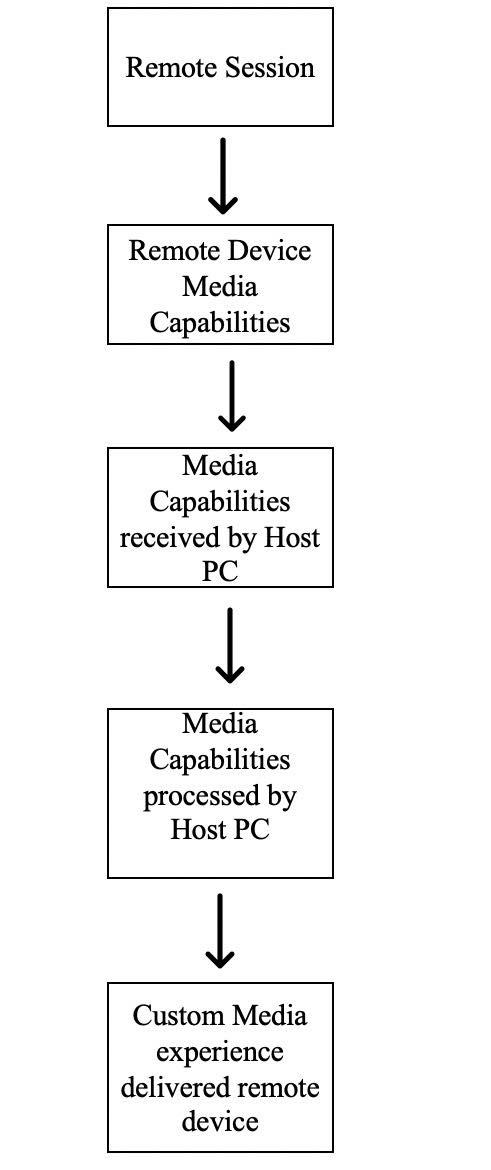}
            \caption{Remote device media capabilities}
            \label{fig:Remote device media capabilities}
\end{figure}

\begin{figure}[h!]
            \centering
            \includegraphics[width=7cm]{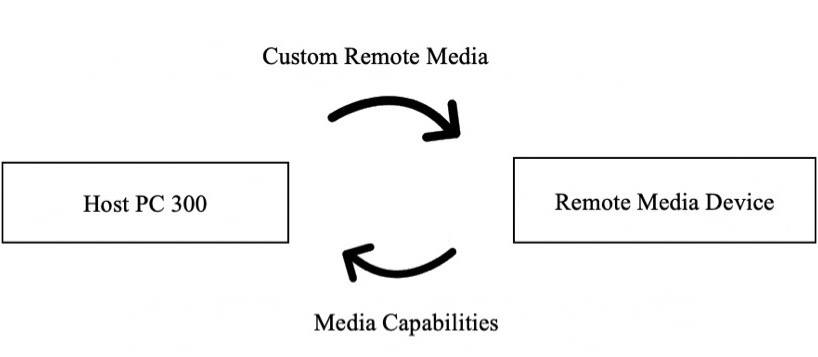}
            \caption{Remote device media capabilities}
            \label{fig:Remote device media capabilities}
\end{figure}

\begin{enumerate}
\item Hardware equipment is easier to develop and manufacture thanks to innovations like 3-D printers or Arduino.
\item The cost of equipment is dropping significantly on account of the mass adoption of consumer gadgets like cell phones.
\item Technologies like sensors or WiFi chips turn into a generally accessible and cheap commodity and are easy to integrate.
\item As the equipment is becoming easier to develop and manufacture, the focus will move far from innovation and will concentrate more on design or problem solving, the same way it happened in software thanks to the growth of APIs.
\item Software today is built with fifty years of oblivious assumptions of a work tool as a primary concern. There are innovators, particularly from the design world, who get through this presumption and make new UIs \cite{senic_2015}.
\end{enumerate}

The ideal system designed should consist of the following characteristics: 
\begin{enumerate}
\item Decentralized: The user interfaces like the light switch shifted onto the smart phone and will now shift away again into smart light switches, speech, or completely new forms like eye tracking. 
\item Specific: Interfaces will move far from a nonexclusive screen towards increasingly explicit interfaces that complete only a few things and that are explicitly intended for that utilization case. This implies explicit interfaces for designers that have attention on haptics, interfaces for elderly people that have an emphasis on straightforwardness and unambiguity, or interfaces for children will have an emphasis on playfulness.
\item Human-centered: Graphical UIs have numerous limitations. They are not accessible to visually impaired or disabled people. They utilize the visual sense and a reduced version of haptics. There can be straining to affect our hand, neck or eyes. Future interfaces will be designed with human science and psychology in mind. It will incorporate more of our human senses.
\item Instant: Putting numerous applications on one device implies that the user will need to deal with menus. With decentralized, explicit interfaces, this will be obsolete. Things will be instant again; the question is not whether an action takes 1, 3, or 5 stages. The question will be if an activity should be possible in a split second or not. This additionally diminishes our cognitive load, which enables us to concentrate on the task at hand or the person in front of us.
\item Simple: Future interfaces will disregard the assumed integration with graphical UI and will concentrate on making things less difficult than existing arrangements.
\item Augmented and virtual: The digital and physical will mix together. Whether through augmented reality glasses or not, the user should have the capacity to peruse setting data about a broken device, not through a cell phone but rather specifically in the surrounding "space" of the object.
\item Passive: An action of a device should be fed as an input to the other. Passive devices are already a major trend in HCI. The classic example of such an application would be turning on the AC of your home when the headlights are approaching the garage \cite{senic_2015}. 
\end{enumerate}

These are few features that a multimodal HCI device must comprise to become successful in the market and accepted by the masses. In the recent past, there are several attempts by companies like which failed deliberately. The Google Glass is one such fine example which became a burden on the user and failed miserably, and the users would not able to leverage the functionalities despite investing a huge chunk of money. Another big factor that also plays a vital role in a device being accepted by a wide range of people is its "cost".

\begin{figure}[h!]
            \centering
            \includegraphics[width=0.5\textwidth]{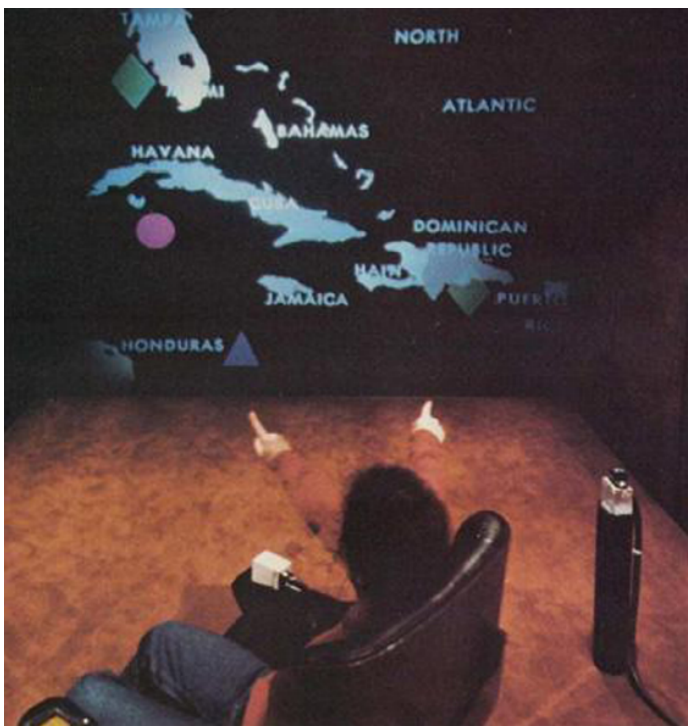}
            \caption{Bolt’s ‘‘Put That There’’ System (Bolt, 1980).}
            \label{fig:Put_That_There}
\end{figure}

\section{Conclusion and Future Work}
With the classification of multimodal systems, there is an ample amount of space for improvement regarding precision, quality, and robustness. The implementation of the multimodal system using supervised learning has high precision and can capture various modalities. The system, designed recently in 2018 and 2019 has advanced in exploring different modalities and its applications, including their application from surgery to find the texture of the cloth and fixing the screw using the robotic arm. This study enables us to develop a system that can implement multimodal fusion using voice, gesture and peripheral device as an input. In most of the systems with fusion, we have analyzed has a huge scope of improvement. In our future work, we plan to develop a system that captures raw EMG data using human limbs along with voice, including input from a third modality, which could be a lever or a keyboard input, that allows to generate input by combination of modalities if the robot misses a gesture or a speech. The initial phase has been implemented and incorporates two modalities, speech and gesture \cite{mohd2017multi}.  The results show significant improvement in comparison with individual modality, with the average error rate reduced to 5.2 percent. Age plays an important factor when using multimodality. Younger people are more comfortable with MYO arm band and other peripheral devices devices , whereas older adults prefer devices with key input in comparison to haptic and gestures. 

Moreover, the voice input modality is still preferable, but it brings another challenge in that it does not accept all the accents of various ethnicities, and the shaky voices of older adults. An incorrect input of voice commands may lead to disaster in situations involving driverless cars. These error can be avoided by training the Artificial Intelligence model using some balanced precentage of data from these subjects or users. A multimodal autonomous car should accept canned pre-defined input and discard others to function properly and seek minimum human intervention. Otherwise, if a car analyzes human speech and receives wrong input, it could be fatal for the users. Likewise, giant robots used in automation or manufacturing units, if developed with multimodal functionality, must understand canned inputs and discard others to function properly. The assisted robot for the blind project is one of the cutting-edge projects in the field of multimodal inputs, but if the system is not efficient enough to adapt, the environment will fail during the evacuation of a building when developing systems for differently abled people,  efficiency and accuracy would be considered the most important criteria. Otherwise, this robotic system might prove fatal to humans. There is a need to develop a device of multimodality with fusion that can be used in myriad industries. During the review, we did not come across a system that can accept multimodal inputs with fusion and is efficient enough to perform tasks in an industrial or a health care sector. So far, the systems designed are either multimodal without fusion, or they accept pre-defined inputs that work under certain conditions.

Moreover, in an industrial environment, it is necessary to move a robot dynamically in any direction to move heavy objects from one place to another. Usually, assembly lines can perform predefined static tasks. Along the similar lines, a speech assistant with an intelligent robotic arm is needed to understand human speech irrespective of ethnicity and pronunciation and that can guide a differently abled person to move in and around a city without the help of other human beings. Google and Microsoft developed speech APIs, which are paid and too expensive to be used by everyday people, and they also have the challenge of understanding of human speech with high accuracy. An error in understanding speech could lead to disaster and may endanger human lives. The ECOMODE device discussed earlier brings a lot of challenges to be used by senior citizens. Emotions and gestures are an easy way to capture input by avoiding speech.

According to our study, there are still avenues open for research in multimodal fusion while exploring different combinations of input modalities. If a system is designed along similar lines, it not only improves the capability of handling industrial robots but also makes life easier for the differently abled.

\bibliographystyle{ieeetr}
\bibliography{references}

\end{document}